\documentclass[a4paper,11pt]{article}
\pdfoutput=1 

\usepackage{jheppub} 

\usepackage[T1]{fontenc} 
\usepackage{bm}
\usepackage{graphicx}
\usepackage{caption}
\usepackage{subcaption}

\newcommand{\REF}{ref.}
\newcommand{\REFS}{refs.}
\newcommand{\FIG}{figure}
\newcommand{\SEC}{section}
\newcommand{\EQ}{eq.}

\title{\boldmath 
Reducing cosmological small scale structure via a large dark matter-neutrino interaction: constraints and consequences}

\author{Bridget Bertoni,}
\author{Seyda Ipek,}
\author{David McKeen}
\author{and Ann E. Nelson}

\affiliation{Department of Physics, University of Washington, Seattle, 
Washington 98195, USA}

\emailAdd{bbertoni@uw.edu}
\emailAdd{ipek@uw.edu}
\emailAdd{dmckeen@uw.edu}
\emailAdd{aenelson@uw.edu}

\abstract{
Cold dark matter explains a wide range of data on cosmological scales. However, there has been a steady accumulation of evidence for discrepancies between simulations and observations at scales smaller than galaxy clusters. Solutions to these small scale structure problems may indicate that simulations need to improve how they include feedback from baryonic matter, or may imply that dark matter properties differ from the standard cold, noninteracting scenario. One promising way to affect structure formation on small scales is a relatively strong coupling of dark matter to neutrinos. We construct an experimentally viable, simple, renormalizable, model with new interactions between neutrinos and dark matter. We show that addressing the small scale structure problems requires dark matter with a mass that is tens of MeV, and a present-day density determined by an initial particle-antiparticle asymmetry in the dark sector. Generating a sufficiently large dark matter-neutrino coupling requires a new heavy neutrino with a mass around 100 MeV. The heavy neutrino is mostly sterile but has a substantial $\tau$ neutrino component, while the three nearly massless neutrinos are partly sterile. We provide the first discussion of how such dark matter-neutrino interactions affect neutrino (especially $\tau$ neutrino) phenomenology. This model can be tested by future astrophysical, particle physics, and neutrino oscillation data. A feature in the neutrino energy spectrum and flavor content from a future nearby supernova would provide strong evidence of neutrino-dark matter interactions. Promising signatures include anomalous matter effects in neutrino oscillations due to nonstandard interactions and a component of the $\tau$ neutrino with mass around 100 MeV.
}

\begin{document} 
\maketitle
\flushbottom


\section{Introduction: Structure Formation at Smaller Scales and Neutrino Interacting Dark Matter}
\label{sec:intro}
The existence of dark matter (DM) is required by a number of experimental results, from galactic rotation curves, to gravitational lensing observations, to the cosmic microwave background (CMB). Because of its success in accounting for data at cosmological scales, the cold DM (CDM) paradigm has become the baseline scenario for studying DM-related physics. Despite these successes, evidence for puzzles at galactic or smaller scales that CDM cannot account for have been accumulating in recent years~\cite{Moore:1999nt,Kauffmann:1993gv,Moore:1994yx,Flores:1994gz, Klypin:1999uc, BoylanKolchin:2011de,Walker:2012td}. These puzzles go by several names: the "missing satellites," "too big to fail," and "core vs. cusp" problems. For general, recent reviews of the small scale structure discrepancies see~\cite{Weinberg:2013aya,Brooks:2014qya}.

Structure formation involves a competition between gravity, which causes density inhomogeneities to grow, and pressure, which resists the gravitational collapse. To examine the scales on which DM structure can form, we must track its history. In the early Universe when the temperature was extremely high, the DM was in 
 thermal equilibrium with the relativistic plasma composed of photons, neutrinos, and possibly other Standard Model (SM) states, via nongravitational interactions that are typically assumed to be present. After the temperature drops below the DM mass, the DM abundance eventually falls out of chemical equilibrium, fixing its (comoving) number density. However, the DM remains in kinetic equilibrium with the plasma via elastic scattering for a longer period of time. These interactions with the relativistic plasma allow the DM to feel a pressure that resists gravitational collapse and therefore suppresses the formation of structure.

As the temperature continues to drop, the DM goes out of kinetic 
equilibrium. This happens at a decoupling temperature $T_{\rm d}$ which can be roughly estimated by determining when the rate for the DM momentum to appreciably change via scattering falls below the expansion rate of the Universe. Since it is nonrelativistic, while in kinetic equilibrium the DM has momentum of order $\sqrt{m_\chi T}$ where $m_\chi$ is its mass and $T$ is the temperature. After $N$ scatterings on the components of the relativistic plasma (which carry momentum $T$), the change in DM momentum is typically about $\sqrt{N}T$. For this change to be comparable to the DM momentum itself implies that $N\sim m_\chi/T$. The rate for $N$ scatterings is $\Gamma=n_r \sigma/N\sim (T\,n_r \sigma)/m_\chi$ where $n_r\propto T^3$ is the plasma (radiation) number density and $\sigma$ is the cross section for scattering. The cross section for scattering on relativistic plasma scales as $\sigma=T^2/\Lambda^4$ where $\Lambda$ is the scale of the operator mediating the interaction. The decoupling temperature is found when $\Gamma=H\sim T^2/M_{\rm Pl}$ where $H$ is the Hubble rate and $M_{\rm Pl}$ is the Planck mass. Solving for the decoupling temperature gives the scaling $T_{\rm d}\sim\Lambda(m_\chi/M_{\rm Pl})^{1/4}$. This expression captures the intuitive expectation that as the interaction strength increases ($\Lambda$ decreases) $T_{\rm d}$ decreases.

Because the Universe is expanding after an initial period of inflation, density perturbations on smaller scales enter the horizon before those of larger size. Since only density perturbations with a size smaller than the horizon can grow, perturbations on smaller scales begin to grow before perturbations on larger scales, so that the decoupling temperature sets a minimum size for DM structures that can form. The growth of DM perturbations on scales smaller than the horizon at $T>T_{\rm d}$ is suppressed by the finite pressure of the coupled DM-plasma gas. For $T<T_{\rm d}$, the DM pressure drops to zero and density perturbations on scales of order the horizon size at $T=T_{\rm d}$ and smaller can grow. Because the (comoving) DM number density remains a constant, and since we know the present DM mass density, the lower bound on the size of unsuppressed DM structures can be expressed as a lower bound on the mass of gravitationally bound DM objects, $M_{\rm cutoff}$. As we will see in \SEC~\ref{sec:sss}, this cutoff can be related to the decoupling temperature via $M_{\rm cutoff}\sim 10^8 M_\odot ({\rm keV}/T_{\rm d})^3$, where $M_\odot\simeq2\times10^{30}~{\rm kg}$ is the mass of the Sun. The standard weakly interacting massive particle (WIMP) CDM scenario with $\Lambda\sim m_\chi\sim 100~\rm GeV$ leads to a decoupling temperature around $10~{\rm MeV}$ and hence $M_{\rm cutoff}\ll M_\odot$, which is too small to be relevant for the small scale structure puzzles~\cite{Green:2005fa,Loeb:2005pm,Bringmann:2009vf}. 

At temperatures below $T_{\rm d}$, the prevailing paradigm, as borne out by DM N-body simulations, is of hierarchical structure formation in which smaller scale DM density perturbations give rise to smaller clumps of matter that merge to form progressively larger objects. The cores of some of the merging clumps of matter survive, resulting in dense, gravitationally bound clumps (called subhalos) within a larger gravitationally bound structure. Simulations based on CDM predict that there should be hundreds to thousands of DM subhalos in a Milky Way size galaxy, in contrast to the few dozen known Milky Way satellite galaxies (for a review see~\cite{2010arXiv1009.4505B}). This discrepancy is termed the "missing satellites" problem. 

The deficit in the number of satellite galaxies could be due to conventional astrophysics, such as the inability of low mass DM subhalos to form stars (for example, due to supernova feedback or reionization blowing regular matter out of subhalos), making them hard to detect (see~\cite{Font:2011ds} and references therein), or it could be due to the properties of the DM particle.  Several ways to change the vanilla CDM paradigm have been suggested  in order to affect structure on small scales and suppress the formation of small DM subhalos, including: (i) the DM could be warmer~\cite{Dodelson:1993je,Colombi:1995ze}, (ii) the DM  could self-interact~\cite{Spergel:1999mh,Tulin:2012wi,Tulin:2013teo,Kaplinghat:2013kqa}, or (iii)   the DM could stay in thermal equilibrium with radiation to lower temperatures than typically expected. To realize option (iii), the DM must interact strongly with the components of the relativistic plasma, either photons, neutrinos, or dark radiation. In this paper we focus on stronger-than-expected interactions of DM with neutrinos, which several groups have considered~\cite{Aarssen:2012fx,Shoemaker:2013tda,Ko:2014bka,Archidiacono:2014nda,Cherry:2014xra}.

Strong neutrino-DM interactions lower $T_{\rm d}$ which increases $M_{\rm cutoff}$, offering a solution to the missing satellites problem by suppressing the formation of smaller halos. Of course, $M_{\rm cutoff}$ must be chosen to be consistent with observations of halo masses.  An analysis of satellite galaxies indicates that the mass of their surrounding DM halos before accretion onto the Milky Way was around $M_{\rm halo} \sim 10^9 M_{\odot}$~\cite{Strigari:2008ib}.  Measurements of the Lyman-$\alpha$ absorption lines in the spectra of distant quasars due to the presence of clumps of intergalactic neutral hydrogen (the ``Lyman-$\alpha$ forest'') indicate that the halos with $M_{\rm halo} \sim 3 \times 10^8 M_{\odot}$ exist~\cite{Viel:2013fqw}.  Similarly, DM substructure can be observed using gravitational lensing, with the smallest structures observed having $M_{\rm halo} \sim 1 \times 10^8 M_{\odot}$~\cite{Vegetti:2012mc,Vegetti:2014lqa}.  Since tidal disruption could cause these observed halo masses to be smaller than the original halo, and since a value for $M_{\rm cutoff} < M_{\rm halo}$ is certainly allowed, in this work we consider models with  $M_{\rm cutoff}$ in the range $10^7 M_{\odot} - 10^9 M_{\odot}$.
This range  requires a decoupling temperature $T_{\rm d}\sim\rm keV$. The scaling of $T_{\rm d}$  implies $(\Lambda^4 m_\chi)^{1/5}\sim 50~{\rm MeV}$. Therefore, in this scenario   the  DM     mass is indicated to be  of order tens of $\rm MeV$. 

There are two other  small scale structure problems in the CDM paradigm, the "core vs. cusp" and "too big to fail" problems.  Simulations of standard CDM predict that the DM density profile $\rho$ in a galaxy should form a cusp at the center, $\rho \propto r^{-1}$~\cite{Navarro:1996gj}. Observations in some dwarf galaxies indicate that actual DM density profiles appear to be more cored, with a constant DM density in the center~\cite{Oh:2010ea}. This is the "core vs. cusp" problem. The "too big to fail problem" is that simulations predict that the most massive Milky Way satellite galaxies should be more massive than they are observed to be. In this paper we will spend most of our time detailing the impact of strong neutrino-DM interactions on the missing satellites problem, but we address the other two problems briefly.

The outline of this paper is as follows. In \SEC~\ref{sec:model}, we build a model of DM  that interacts strongly enough with neutrinos to obtain a cutoff mass in the range $10^7 M_{\odot} - 10^9 M_{\odot}$ and examine the constraints on it. Section~\ref{sec:sss} contains a detailed calculation of the cutoff mass in the model. Effects on supernovae are especially interesting and we examine them in this model in \SEC~\ref{sec:sn}. In \SEC~\ref{sec:future} we discuss some possible tests of the model and we conclude in \SEC~\ref{sec:conc}.


\section{The model}
\label{sec:model}
\subsection{Ingredients and basics}
Interactions between neutrinos and SM gauge singlets, such as DM, can be safely 
generated through the "neutrino portal." In this scenario, couplings of the SM 
to DM occur through the operator $H\ell$, where $H$ is the Higgs 
doublet and $\ell$ is a lepton doublet containing a neutrino and a charged 
lepton. An effective 4-fermi 
interaction between neutrinos and DM can be generated that looks schematically like $\left(H\ell\right)^2\left(\rm DM\right)^2$. At the renormalizable level, this higher dimensional operator arises due to the exchange of a 
mediator that is either neutral or charged under the symmetry that 
is typically invoked to keep the DM stable. If the mediator is neutral, then 
exchange of this mediator also leads to neutrino and DM self-interactions. 
To focus   primarily on  DM-neutrino interactions, we 
study the case where the mediator is also charged under the DM stabilization 
symmetry.

In light of the discussion above, we introduce a complex scalar, $\phi$, and 
a Dirac fermion, $\chi$, which are oppositely charged under a global, conserved 
U(1)$_{\rm d}$ that acts as the DM stabilization symmetry. 
The SM fields are all neutral with respect to this U(1)$_{\rm d}$. 
The lighter of $\phi$ and $\chi$ is therefore stable 
and is our DM candidate.  
For definiteness, and because the opposite situation gives qualitatively the same results, we focus on the situation where 
the DM is fermionic with $\chi$ lighter than $\phi$.

Additionally, we give $\phi$ lepton number 
$-1$ so that we can generate an effective 
DM-neutrino coupling through the operator $\phi\bar\chi\nu$ without 
breaking lepton number.
The other ingredients in the model are a pair of left-handed Weyl fermions, 
$N_{1,2}$, with lepton number $-1$ and $+1$ respectively, that are SM gauge 
singlets, i.e.\ sterile neutrinos. 

We assume that lepton number is conserved in interactions involving $N_{1,2}$. The observed masses of the light neutrinos could be of the lepton-number violating Majorana type,  arising from
  other lepton-number--violating interactions at a high scale, or  Dirac. Although Dirac neutrino masses can easily be made consistent with our model, for definiteness we will assume the tiny observed masses are Majorana, arising  e.g.\ through a standard seesaw scenario. Below the seesaw scale, 
the terms in the Lagrangian relevant for the neutrino masses are given by
\begin{align}
-{\cal L}_{\rm m}&=\frac{m_{ij}}{\langle H\rangle^2}H\ell_i H\ell_j+MN_1 N_2+\lambda_iN_1 H\ell_i+{\rm h.c.},
\label{eq:Lm}
\end{align}
where $i,j=e,\mu,\tau$ are lepton flavor indices and $H$ is the Higgs doublet. Electroweak and 
Lorentz indices have been suppressed. $m_{ij}$ is the effective Majorana mass matrix for the active neutrinos which can be generated at a very high scale by interactions that violate lepton number, the details of which are irrelevant for us. We assume that each of the entries in $m$ is much smaller than $M$.

The interaction of the sterile neutrinos with the DM and mediator is given by
\begin{align}
-{\cal L}_{\rm int}&=\left(y_1\phi^\ast N_1+y_2\phi N_2\right)\chi_L +{\rm h.c.},
\label{eq:LintF}
\end{align}
We have assumed that the couplings of the right-handed component of $\chi$ can be ignored compared to those of the left-handed component---reversing or relaxing this assumption does not change any of the physics we are interested in.

After electroweak symmetry is broken, the Higgs field gets a 
vacuum expectation value, $\langle H\rangle\equiv v=174~\rm GeV$, which leads, in the basis $(\nu_i,N_1^\ast,N_2)$, to the neutrino mass matrix,
\begin{align}
\left(  \begin{array}{ccc}
    m_{ij} & \lambda_j v & 0 \\ 
    \lambda_i v & 0 & M \\
    0 & M & 0 \\
  \end{array}\right).
\end{align}
$N_1^\ast$ pairs up with
\begin{align}
\hat\nu_4=\frac{MN_2+\sum_i\lambda_i v \nu_i}{\sqrt{M^2+\sum_i\lambda_i^2 v^2}}
\end{align}
to form a Dirac fermion $\hat N=\left(\hat\nu_4,N_1^\ast\right)^{\rm T}$ with mass $m_4=\sqrt{M^2+\sum_i\lambda_i^2 v^2}$. To avoid limits on the number of neutrino species present during the time of neutrino decoupling from measurements of the CMB~\cite{Ade:2013zuv} as well as from big bang nucleosynthesis (BBN)~\cite{Iocco:2008va}, we take $m_4>10~\rm MeV$. The orthogonal linear combinations of $\nu_i$ and $N_2$ furnish three Majorana neutrinos with masses $m_{1,2,3}$ which are extremely small compared to $m_4$, at most ${\cal O}\left(0.5~\rm eV\right)$. We write the relationship between the mass eigenstates and the flavor eigenstates explicitly using the unitary matrix $U$ that diagonalizes the mass matrix,
\begin{align}
\nu_i=U_{ij}\hat\nu_j,
\end{align}
with $i=e,\mu,\tau,N$ (defining $\nu_N\equiv N_2$) and $j=1,\dots,4$. 

The mediator $\phi$ decays to $\bar\chi$ and antineutrinos through \EQ~\eqref{eq:LintF}. The rate for this is
\begin{align}
\Gamma_{\phi\to\bar\nu\bar\chi}=\sum_i\left|U_{Ni}\right|^2\frac{y_2^2m_\phi}{16\pi},
\label{eq:GammaPhi}
\end{align}
where the sum runs over kinematically allowed neutrinos, and we neglect the light neutrino masses and $m_\chi$.
We will be most interested in the case where the heavy neutrino can decay invisibly to $\chi\bar\chi$ plus a light neutrino through an 
intermediate $\phi$ with a rate,
\begin{align}
\Gamma_{\hat N\to\nu\chi\bar\chi}=\frac{\left(y_1^2+\left|U_{N4}\right|^2y_2^2\right)m_4}{32\pi}\times\begin{cases} 1&\mbox{if } m_\phi \ll m_4 \\ 
\left(1-\left|U_{N4}\right|^2\right)\frac{y_2^2}{192\pi^2}\left(\frac{m_4}{m_\phi}\right)^4 & \mbox{if } m_\phi \gg m_4, \end{cases}
\label{eq:GammaNu4}
\end{align}
where we have ignored $m_\chi$. As long as this is kinematically allowed ($m_4>2m_\chi$, since the light neutrino masses are negligible) it is the dominant decay channel for the heavy neutrino. The heavy neutrino can also decay visibly through the weak neutral current. The rate for the decay to $\nu e^+ e^-$, for example, is
\begin{align}
\Gamma_{\hat N\to\nu e^+ e^-}= \left(1-\left|U_{N4}\right|^2\right)\frac{G_F^2 m_4^5}{192\pi^3}.
\end{align}
In the phenomenologically interesting region $G_F m_4^2,G_F m_\phi^2\ll 1$, so the visible decays of the heavy neutrino are highly suppressed.

\begin{figure}[tbp]
\centering
\includegraphics[width=.45\textwidth]{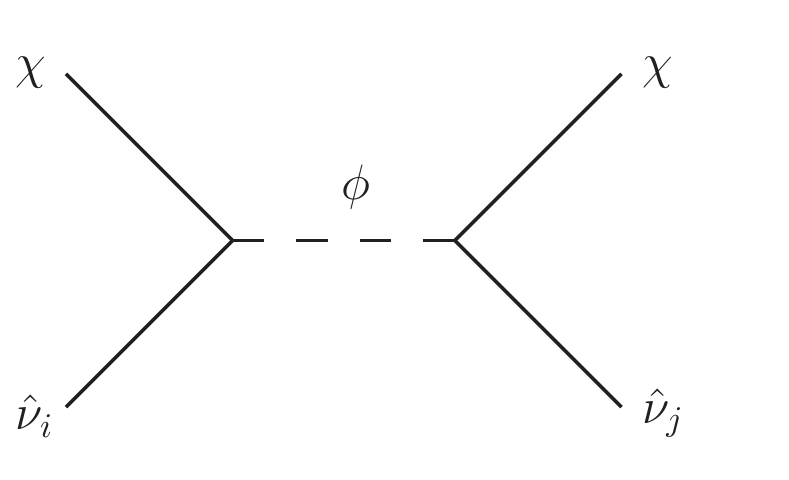}
\caption{\label{fig:diagram} Diagram relevant for neutrino scattering on DM and for DM annihilation to neutrinos.}
\end{figure}

We will be particularly interested in the cross section for the light neutrinos to scatter on DM at rest, through diagrams like that shown in \FIG~\ref{fig:diagram}. Defining $\sigma_{\hat\nu_i\chi}$ as the cross section for the $i$th neutrino mass eigenstate to scatter, $\hat\nu_i\chi\to\sum_{j=1}^3\hat\nu_j\chi$, we have
\begin{align}
\sigma_{\hat\nu_i\chi}&=\frac{\left|U_{Ni}\right|^2}{\left|U_{e4}\right|^2+\left|U_{\mu 4}\right|^2+\left|U_{\tau 4}\right|^2}\sigma_{\nu\chi},\quad\sigma_{\nu\chi}=\sum_{i=1}^3 \sigma_{\hat\nu_i\chi},
\label{eq:sigmai}
\end{align}
with
\begin{equation}
\begin{aligned}
\frac{d\sigma_{\nu\chi}}{dE_\nu^\prime}=\frac{g^4}{32\pi}m_\chi&\left\{\frac{1}{\left(m_\phi^2-m_\chi^2-2m_\chi E_\nu\right)^2+m_\phi^2\Gamma_\phi^2}\right.
\\
&\quad\quad\quad\quad\left.+\frac{{E_\nu^\prime}^2/E_\nu^2}{\left(m_\phi^2-m_\chi^2+2m_\chi E_\nu^\prime\right)^2+m_\phi^2\Gamma_\phi^2}\right\}.
\end{aligned}
\end{equation}
In this expression, $E_\nu$ is the initial neutrino energy and $E_\nu^\prime$ is the final neutrino energy  which is in the range $E_\nu/\left(1+2E_\nu/m_\chi\right)<E_\nu^\prime<E_\nu$. We have ignored the light neutrino masses, made use of the unitarity of $U$, and defined the coupling
\begin{align}
g\equiv y_2\sqrt{\left|U_{e4}\right|^2+\left|U_{\mu 4}\right|^2+\left|U_{\tau 4}\right|^2}.
\end{align}
Without loss of generality, we will set $g>0$ throughout this paper.

In the limit that the neutrino energy is small, $E_\nu\ll m_{\chi,\phi}$, the cross section becomes
\begin{align}
\sigma_{\nu\chi}&=\frac{g^4}{8\pi}\frac{E_\nu^2}{\left(m_\phi^2-m_\chi^2\right)^2}
=5\times10^{-38}{\rm cm}^2\left(\frac{g}{0.3}\right)^4\left(\frac{E_\nu}{1~\rm keV}\right)^2\left(\frac{40~\rm MeV}{m_\phi}\right)^4,
\end{align}
ignoring terms of order $m_\chi^2/m_\phi^2$ on the RHS. This form for the cross section matches on to $\sigma=T^2/\Lambda^4$ with $\Lambda\sim\sqrt{m_\phi^2-m_\chi^2}/g$. As mentioned in the introduction, to have $10^7M_\odot\lesssim M_{\rm cutoff}\lesssim10^9M_\odot$, we require that $\Lambda$ and $m_\chi$ are  ${\cal O}\left({\rm few}\times 10~{\rm MeV}\right)$. In other words, we need $m_\chi$ and $m_\phi$ to be tens of MeV and $g\gtrsim 0.1$.

When the temperature of the Universe is larger than $m_\chi$, the DM and anti-DM will exist in chemical equilibrium with the rest of the constituents of the plasma. As the Universe cools below $m_\chi$, the DM and anti-DM number densities are depleted through annihilation to light neutrinos via $\phi$ exchange (also through diagrams like the one in \FIG~\ref{fig:diagram}) which occurs with a cross section
\begin{align}
\label{eq:ann_xsec}
\sigma_{\rm ann}v&=\frac{g^4m_\chi^2}{16\pi m_\phi^4}=3\times10^{-20}\frac{{\rm cm}^3}{\rm s}\left(\frac{g}{0.3}\right)^4\left(\frac{m_\chi}{20~\rm MeV}\right)^2\left(\frac{40~\rm MeV}{m_\phi}\right)^4.
\end{align}
This process sets a lower limit on the DM mass to avoid the production of neutrinos during BBN. The requirement is that $m_\chi\gtrsim 10~\rm MeV$~\cite{Iocco:2008va,Serpico:2004nm}. Additionally, for parameter values motivated by small scale structure considerations, this annihilation cross section is too large for the DM to be 
a thermal relic. We assume that its relic density is set by a primordial 
DM--anti-DM asymmetry. Whether this asymmetry is connected to the baryon 
asymmetry is beyond the scope of this work.

There are also constraints on the strength of the DM-neutrino interaction from measurements of the Lyman-$\alpha$ forest~\cite{Wilkinson:2014ksa} as well as the CMB~\cite{Boehm:2013jpa,Nollett:2014lwa}, which again imply that the DM has a mass greater than about 10~MeV. 

To determine whether $g\gtrsim0.1$ and $m_{\chi,\phi}\sim{\rm few}\times 10~\rm MeV$ are feasible requires examining the constraints on $\left|U_{e4}\right|$, $\left|U_{\mu 4}\right|$, and $\left|U_{\tau 4}\right|$. We do this in the next section. We assume that $m_4>2m_\chi$, so that the heavy neutrino decays invisibly.


\subsection{Neutrino mixing matrix elements}
Below, we determine what values of the elements of the neutrino mixing matrix, $U$, are allowed by data from lepton and meson decays and neutrino oscillation experiments. We pay particular attention to $\left|U_{e4}\right|$, $\left|U_{\mu 4}\right|$, and $\left|U_{\tau 4}\right|$ since they directly enter the cross section relevant for keeping DM in thermal equilibrium with the neutrinos. Limits on these elements are summarized in \FIG~\ref{fig:Ulimit}. We defer discussion of constraints from supernovae to \SEC~\ref{sec:sn}.

\subsubsection{Limits on $\left|U_{e4}\right|$, $\left|U_{\mu 4}\right|$, and $\left|U_{\tau 4}\right|$ from particle decays}
\label{sec:Ulimits}
We now examine   the existing limits on $\left|U_{e4}\right|$, $\left|U_{\mu 4}\right|$, and $\left|U_{\tau 4}\right|$ in the case of an invisibly decaying heavy neutrino with mass above around $10~\rm MeV$ that can be derived from meson and lepton decays. 

\begin{figure}[tbp]
\centering
\includegraphics[width=.65\textwidth]{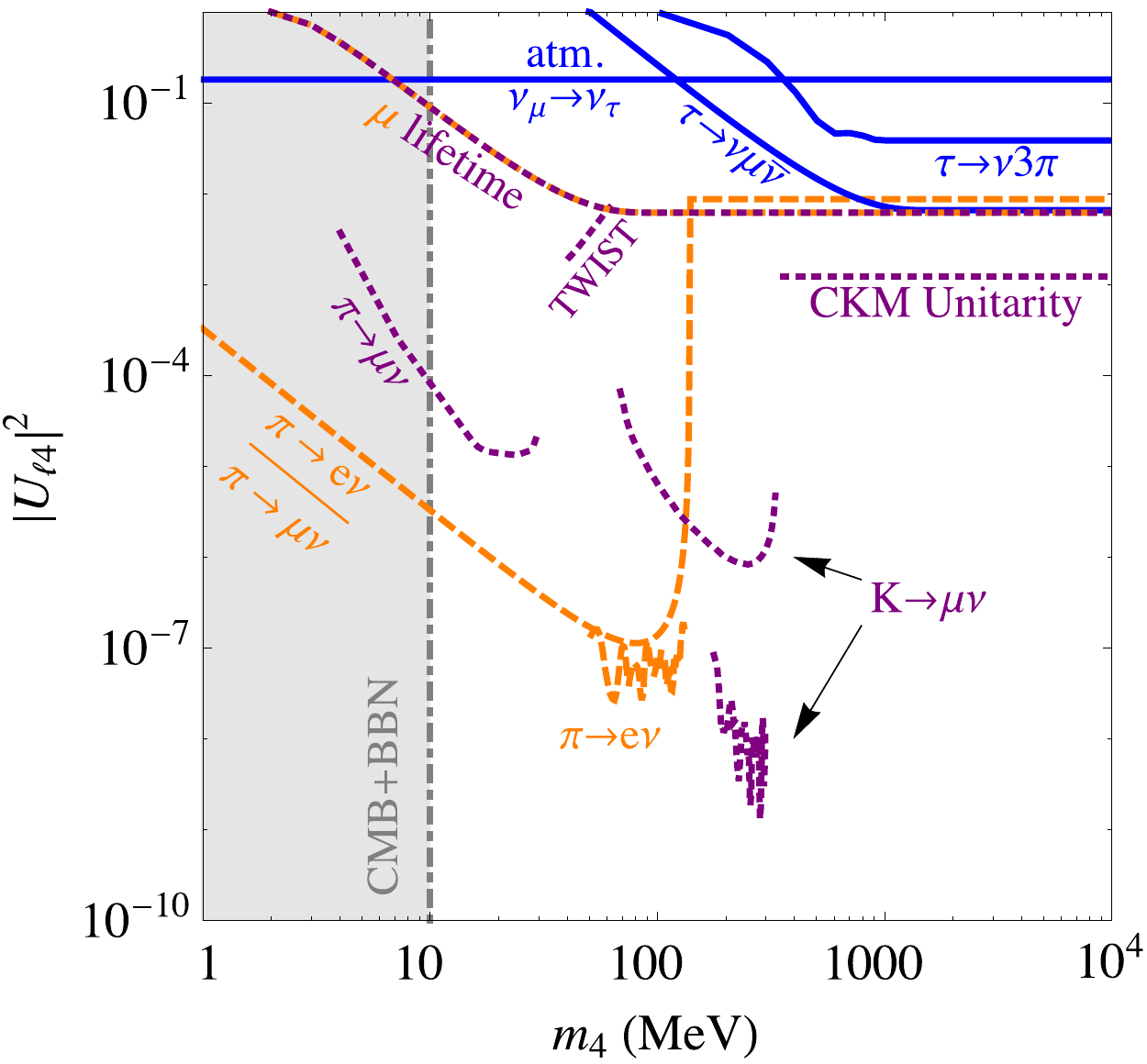}
\caption{\label{fig:Ulimit} 90\% C.\,L. upper limits on $\left|U_{\ell4}\right|^2$, $\ell=e,\mu,\tau$, as functions of the heavy neutrino mass $m_4$ from meson and lepton decays. Limits on $\left|U_{e4}\right|^2$ (dashed, orange) come from searches in $\pi\to e\nu$~\cite{Britton:1992xv}, the ratio of $\Gamma(\pi\to e\nu)/\Gamma(\pi\to\mu\nu)$~\cite{Britton:1992pg,Czapek:1993kc,Marciano:1993sh} and the measurement of the muon lifetime~\cite{Tishchenko:2012ie}. $\left|U_{\mu 4}\right|^2$ limits (dotted, purple) are derived from peak searches in $\pi\to \mu\nu$~\cite{Abela:1981nf} and $K\to \mu\nu$~\cite{Hayano:1982wu,Artamonov:2014eqa}, the muon lifetime measurement~\cite{Tishchenko:2012ie}, the energy spectrum in muon decay (labeled "TWIST")~\cite{Bayes:2010fg,Gninenko:2010pr}, and the unitarity of the measurements of the CKM matrix~\cite{Agashe:2014kda}. We have derived limits on $\left|U_{\tau 4}\right|^2$ (blue, solid) from the rate for $\tau\to\mu\nu\bar\nu$ and data from \REF~\cite{Barate:1997zg} on $\tau\to\nu3\pi$. We also show the limit on $\left|U_{\tau 4}\right|$ from atmospheric neutrino oscillations described in \SEC~\ref{sec:osc} and the lower bound on $m_4$ from BBN and CMB measurements. Where rates depend on more than one $\left|U_{\ell4}\right|$, we assume only one is dominant to produce each limit. See text for details.}
\end{figure}

We first focus on decays of a meson $M$ to a lepton $\ell$ and neutrino mass eigenstate $\hat\nu_i$, $M^+\to\ell^+\hat\nu_i$. Since the heavy neutrino decays invisibly, all final state neutrino mass eigenstates result in the same signal, $\ell^+$ and missing energy, up to a difference in the energy of the $\ell^+$. The light neutrino masses can all be neglected, while the heavy neutrino mass must be retained. The decay rate to light neutrinos is
\begin{align}
\Gamma_{M^+\to \ell^+\hat\nu_{1,2,3}}&=\sum_{i=1}^3\left|U_{\ell i}\right|^2\Gamma_{M^+\to \ell^+}^{\rm SM}=\left(1-\left|U_{\ell 4}\right|^2\right)\Gamma_{M^+\to \ell^+}^{\rm SM}
\end{align}
and the rate to the heavy neutrino is
\begin{align}
\Gamma_{M^+\to \ell^+\hat\nu_4}&=\left|U_{\ell 4}\right|^2\rho_{M\ell}\left(m_4\right)\Gamma_{M^+\to \ell^+}^{\rm SM}.
\end{align}
$\Gamma_{M^+\to \ell^+}^{\rm SM}$ is the rate for this process as calculated in the SM for a massless neutrino. $\rho_{M\ell}\left(m_4\right)$ is a factor that reflects the reduced phase space available as well as possible enhancement to helicity-suppressed decays with $\rho_{M\ell}\left(m_4=0\right)=1$ and $\rho_{M\ell}\left(m_4\ge m_M-m_\ell\right)=0$. There are therefore two signatures of a heavy neutrino in this decay: (i) a change in the total rate for $M\to \ell$ from the SM expectation and (ii) if kinematically allowed, a peak in the $\ell$ energy spectrum at $(m_M^2+m_\ell^2-m_4^2)/2m_M$ in the $M$ rest frame.

Decays with more than two particles in the final state, such as muon decay, leptonic $\tau$ decays, and semileptonic kaon decays, are modified analogously, with a straightforward adjustment of the phase space and the possible inclusion of a second $\left|U_{\ell 4}\right|$ if two neutrinos are in the final state. We now discuss the available data.

Peak searches in $\pi\to e\nu$~\cite{Britton:1992xv} strongly constrain $\left|U_{e4}\right|$ while searches for peaks in $\pi\to \mu\nu$~\cite{Abela:1981nf}, and $K\to \mu\nu$~\cite{Hayano:1982wu,Artamonov:2014eqa} similarly limit $\left|U_{\mu4}\right|$. We show these limits in \FIG~\ref{fig:Ulimit}.

Limits on $\left|U_{e4}\right|$ and $\left|U_{\mu4}\right|$ can be obtained by comparing $\Gamma_{\pi\to e\nu}$ to $\Gamma_{\pi\to\mu\nu}$. Defining
\begin{align}
R=\left(\frac{\Gamma_{\pi\to e\nu}}{\Gamma_{\pi\to\mu\nu}}\right)_{\rm exp}\Bigg/\left(\frac{\Gamma_{\pi\to e\nu}}{\Gamma_{\pi\to\mu\nu}}\right)_{\rm SM}=\frac{1+\left|U_{e4}\right|^2\left[\rho_{\pi e}\left(m_4\right)-1\right]}{1+\left|U_{\mu4}\right|^2\left[\rho_{\pi\mu}\left(m_4\right)-1\right]},
\end{align}
current data~\cite{Britton:1992pg,Czapek:1993kc} and SM prediction~\cite{Marciano:1993sh} gives $R=0.996\pm 0.003$. This is particularly constraining on $\left|U_{e4}\right|$ since $\pi\to e\hat\nu_4$ is not as helicity-suppressed as the decay to light neutrinos. In \FIG~\ref{fig:Ulimit} we show the limit that $R$ implies on $\left|U_{e4}\right|$ if we assume that $\left|U_{\mu4}\right|=0$ which gives a conservative limit for $m_4<m_\pi$.

Nonzero values of $\left|U_{e4}\right|$ and $\left|U_{\mu4}\right|$ can also affect the relationship between the Fermi constant extracted from measurements of the muon lifetime, $G_\mu$, and determinations using high energy data. In the on-shell renormalization scheme, for example, the Fermi constant can be expressed in terms of the $W$ and $Z$ boson masses and the fine structure constant through
\begin{align}
G_F=\frac{\pi\alpha}{\sqrt2\left(1-m_W^2/m_Z^2\right)m_W^2\left(1-\Delta r\right)},
\label{eq:GFewk}
\end{align}
where $1-\Delta r=0.9636\pm0.0004$ encodes radiative corrections. The presence of a heavy neutrino with $m_4>m_\mu-m_e$ changes the relationship between $G_F$ measured in this way and $G_\mu$ via
\begin{align}
G_\mu^2=\left(1-\left|U_{e4}\right|^2\right)\left(1-\left|U_{\mu 4}\right|^2\right)G_F^2.
\label{eq:Gmu}
\end{align}
The expression when $m_4<m_\mu-m_e$ is more complicated but straightforward. The very precise measurement of the muon lifetime~\cite{Tishchenko:2012ie} gives $G_\mu=\left(1.1663787\pm0.0000006\right)\times10^{-5}~{\rm GeV}^{-2}$. Using $m_W=80.385\pm0.015~{\rm GeV}$ and $m_Z=91.1876\pm0.0021~{\rm GeV}$ results in $G_F=\left(1.168\pm0.001\right)\times10^{-5}~{\rm GeV}^{-2}$. We use these values to set an upper limit on the larger of $\left|U_{e4}\right|$ or $\left|U_{\mu4}\right|$, conservatively assuming that the smaller of the two can be neglected, in \FIG~\ref{fig:Ulimit}. For $m_4>m_\mu-m_e$, the limit is $\left|U_{e4}\right|$, $\left|U_{\mu 4}\right|<7.9\times10^{-2}$ at 90\%~C.\,L.

For $m_4<m_\mu-m_e$, the shape of the $e^+$ energy spectrum in $\mu^+$ decay is modified. This spectrum was most accurately measured in~\cite{Bayes:2010fg} which was used to set a limit on $\left|U_{\mu 4}\right|$ in~\cite{Gninenko:2010pr} for $m_4>40~{\rm MeV}$ which we also show in \FIG~\ref{fig:Ulimit}, labeled as "TWIST."\footnote{The reason we only show the limit for $m_4>40~{\rm MeV}$ is that there is a gap in the limit on $\left|U_{\mu 4}\right|$ between the regions probed by $\pi\to \mu\nu$ and $K\to \mu\nu$, i.e. for $40~{\rm MeV}< m_4 < 80~{\rm MeV}$~\cite{Gninenko:2010pr}.}

There is also a constraint on $\left|U_{\mu4}\right|$ that can be derived using the unitarity of the quark mixing (CKM) matrix, $V$. $V_{ud}$ is most accurately measured using superallowed nuclear beta decays. For $m_4$ larger than several $\rm MeV$, the rates for these are proportional to
\begin{align}
\left(1-\left|U_{e4}\right|^2\right)\left|V_{ud}\right|^2G_F^2.
\end{align}
$V_{ud}$ is extracted by dividing this by $G_\mu^2$. Doing so gives~\cite{Agashe:2014kda}
\begin{align}
\frac{\left|V_{ud}\right|}{\sqrt{1-\left|U_{\mu4}\right|^2}}=0.97425\pm0.00022.
\end{align}
A value of $V_{us}$ can be extracted from $K_L\to\pi^- e^+\nu$ decay. For $m_4>m_{K^0}-m_{\pi^\pm}-m_e$, the rate for this is proportional to
\begin{align}
\left(1-\left|U_{e4}\right|^2\right)\left|V_{us}\right|^2G_F^2
\label{eq:Vud}
\end{align}
and again dividing by $G_\mu^2$ results in~\cite{Agashe:2014kda}
\begin{align}
\frac{\left|V_{us}\right|}{\sqrt{1-\left|U_{\mu4}\right|^2}}=0.2253_{-0.0013}^{+0.0015}.
\label{eq:Vus}
\end{align}
Squaring then adding \eqref{eq:Vud} and \eqref{eq:Vus} and using the unitarity of the CKM matrix implies that
\begin{align}
\frac{1-\left|V_{ub}\right|^2}{1-\left|U_{\mu4}\right|^2}=0.9999_{-0.0007}^{+0.0008}.
\end{align}
At this level $\left|V_{ub}\right|$ is negligible and can be ignored. Doing so, this translates into the constraint $\left|U_{\mu 4}\right|<3.5\times10^{-2}$ at 90\%~C.\,L.

To find limits on $\left|U_{\tau 4}\right|$, we look at processes involving the $\tau$ neutrino such as $\tau$ decays or decays of $D_s$ mesons. Existing searches using $D_s$ decays for heavy neutrinos all look for visible decays of the heavy neutrino, so they are not sensitive to this scenario. Consequently, we focus on $\tau$ decays. In deriving our limits on $\left|U_{\tau 4}\right|$ below, we assume that $\left|U_{e4}\right|$, $\left|U_{\mu 4}\right|\ll\left|U_{\tau 4}\right|$. In this case, the total rate for $\tau$ to decay to a generic final state $X$ plus missing energy is
\begin{align}
\Gamma_{\tau\to X\nu}&=\left[1+\left|U_{\tau4}\right|^2\left(\rho_{\tau X}\left(m_4\right)-1\right)\right]\Gamma_{\tau\to X}^{\rm SM},
\label{eq:taudecay}
\end{align}
where, as before, $\rho_{\tau X}\left(m_4\right)$ is a kinematic factor that depends on the heavy neutrino mass.

Because leptonic $\tau$ decay rates are well-predicted and well-measured, they can offer meaningful constraints on $\left|U_{\tau 4}\right|$. We can use the measurements of the branching ratios for $\tau\to e\bar\nu\nu$ and $\tau\to \mu\bar\nu\nu$ of $17.83\pm0.04\%$ and $17.41\pm0.04\%$ respectively~\cite{Agashe:2014kda} along with the independently measured $\tau$ lifetime, $\left(290.17\pm 0.62\right)\times10^{-15}~\rm s$~\cite{Belous:2013dba}, to determine the experimental rates,
\begin{equation}
\begin{aligned}
\Gamma_{\tau\to e\bar\nu\nu}^{\rm exp}&=\left(4.04\pm0.09\right)\times10^{-13}~{\rm GeV},
\\
\Gamma_{\tau\to\mu\bar\nu\nu}^{\rm exp}&=\left(3.95\pm0.09\right)\times10^{-13}~{\rm GeV}.
\end{aligned}
\end{equation}
Using these with the SM expectations for these rates which take into account the error on $m_\tau$~\cite{Pich:2013lsa},
\begin{equation}
\begin{aligned}
\Gamma_{\tau\to e\bar\nu\nu}^{\rm SM}&=\left(4.031\pm0.001\right)\times10^{-13}~{\rm GeV},
\\
\Gamma_{\tau\to\mu\bar\nu\nu}^{\rm SM}&=\left(3.920\pm0.001\right)\times10^{-13}~{\rm GeV},
\end{aligned}
\end{equation}
and the expression in \eqref{eq:taudecay} can limit $\left|U_{\tau 4}\right|$. The constraint from $\tau\to\mu$ decay, shown in \FIG~\ref{fig:Ulimit},\footnote{Our limits on $\left|U_{\tau 4}\right|$ for a heavy neutrino that decays invisibly differ substantially from those in~\cite{Helo:2011yg} which also considered shifts in leptonic $\tau$ decays. The main reason for this is that we consider a unitary neutrino mixing matrix whereas~\cite{Helo:2011yg} does not [effectively making the replacement $\rho_{\tau X}\left(m_4\right)\to\rho_{\tau X}\left(m_4\right)+1$ in \eqref{eq:taudecay}] with the consequence that our limits weaken as the heavy neutrino mass is decreased. This is to be expected for a unitary mixing matrix since the heavy neutrino becomes indistinguishable from the light neutrinos as it is made lighter. Additionally, since the presence of a heavy neutrino in the final state also affects the other decay modes, we limit the change in the leptonic \emph{rate} itself using the branching ratio and the independent determination of the $\tau$ lifetime. In contrast~\cite{Helo:2011yg} limited shifts of the branching ratio effectively assuming the total width was unchanged.} is stronger since the central value of the measured rate is further from the SM expectation than the $\tau\to e$ mode, although still in agreement. For $m_4>m_\tau-m_\mu$ the 90\%~C.\,L. limit is $\left|U_{\tau 4}\right|<8\times10^{-2}$ and weakens to $\left|U_{\tau 4}\right|<0.4$ at $m_4=130~\rm MeV$.

We can also set a limit on $\left|U_{\tau 4}\right|$ by looking for changes in the differential rates for $\tau$ decays to multiparticle final states. This procedure was undertaken by the ALEPH collaboration~\cite{Barate:1997zg} to place an upper limit of $18.2~\rm MeV$ at 95\% C.\,L. on the mass of the $\tau$ neutrino using $\tau\to\nu3\pi$ and $\tau\to\nu5\pi$ decays. We use the data for the $\tau\to\nu3\pi$ rate as a function of the three pion invariant mass from~\cite{Barate:1997zg}, modeling the $\tau\to\nu3\pi$ decay as occurring through the chain $\tau\to\nu a_1\to\nu\pi\rho\to\nu3\pi$ to set a limit on $\left|U_{\tau 4}\right|$ varying $m_4$, also shown in \FIG~\ref{fig:Ulimit}. This limit is less strong than what we derived from $\tau\to\mu$ decays, partially due to the fact that properly modeling the $3\pi$ rate is nontrivial. Properly modeling the $5\pi$ decay mode is even more difficult but could offer an improvement in the limit due to the reduced phase space available which enhances the effects of a massive neutrino.

As mentioned above, solutions to small scale structure problems imply that $g=y_2\sqrt{\sum_\ell \left|U_{\ell 4}\right|^2}\sim 0.3$. Given the constraints outlined above, and without increasing $y_2$ to nonpertubative values, this is only possible to achieve with $\left|U_{\tau 4}\right|\gtrsim0.1$. In this case $m_4\lesssim 300~\rm MeV$ and $\left|U_{e 4}\right|$, $\left|U_{\mu 4}\right|\ll\left|U_{\tau 4}\right|$. In the rest of the paper, we therefore simplify our analysis by making the approximation that $\lambda_{e,\mu}=0$ in \EQ~\eqref{eq:Lm}. In that case, $\left|U_{e4}\right|=\left|U_{\mu 4}\right|=0$ and $U_{\tau 4}\equiv\sin\theta_\tau$ with $\theta_\tau=\tan^{-1}\left(-\lambda_\tau v/M\right)$. The light Majorana neutrinos are linear combinations of $\nu_e$, $\nu_\mu$ and $\nu_{\tau N}\equiv \cos\theta_\tau \nu_\tau-\sin\theta_\tau N_2$ while the heavy Dirac neutrino has mass $m_4= M/\cos\theta_\tau$ and contains $\hat\nu_4=\cos\theta_\tau N_2+\sin\theta_\tau \nu_\tau$ and $N_1^\ast$.\footnote{The reason that we did not choose to add just a single Weyl sterile neutrino earlier is that in such a scenario, after integrating the sterile neutrino out, the light neutrino mass matrix element $m_{ij}$ receives contributions proportional to the product of mixing angles $\theta_i\theta_j M$, where $M$ is the sterile neutrino mass and the mixing angle between active neutrino $i$ and the sterile neutrino is again $\theta_i\sim\lambda _i v/M$. Obtaining a mixing angle large enough to be interesting in this case requires a sterile neutrino that is too light to avoid cosmological difficulties.}


\subsubsection{Constraints from neutrino oscillation experiments}
\label{sec:osc}
In the limit that $U_{e4}$ and $U_{\mu 4}$ are zero, we can parameterize the $4\times4$ matrix $U$ using only four angles, $\theta_\tau$ as defined above, $\theta_{12}$, $\theta_{13}$, and $\theta_{23}$,
\begin{equation}
\begin{aligned}
U=\left(\begin{array}{cccc}
    1 & 0 & 0 & 0 \\ 
    0 & 1 & 0 & 0 \\
    0 & 0 & c_\theta & s_\theta \\
    0 & 0 & -s_\theta & c_\theta
  \end{array}\right)
  \left(\begin{array}{cccc}
    1 & 0 & 0 & 0 \\ 
    0 & c_{23} & s_{23} & 0 \\
    0 & -s_{23} & c_{23} & 0 \\
    0 & 0 & 0 & 1
  \end{array}\right)
  \left(\begin{array}{cccc}
    c_{13} & 0 & s_{13} & 0 \\ 
    0 & 1 & 0 & 0 \\
    -s_{13} & 0 & c_{13} & 0 \\
    0 & 0 & 0 & 1
  \end{array}\right)
  \left(\begin{array}{cccc}
    c_{12} & s_{12} & 0 & 0 \\
    -s_{12} & c_{12} & 0 & 0 \\
    0 & 0 & 1 & 0 \\
    0 & 0 & 0 & 1
  \end{array}\right)
\\
=\left(\begin{array}{cccc}
    c_{12}c_{13} & c_{13}s_{12} & s_{13} & 0 \\ 
    -c_{23}s_{12}-c_{12}s_{13}s_{23} & c_{12}c_{23}-s_{12}s_{13}s_{23} & c_{13}s_{23} & 0 \\
    -c_\theta\left(c_{12}c_{23}s_{13}-s_{12}s_{23}\right) & -c_\theta\left(c_{23}s_{12}s_{13}+c_{12}s_{23}\right) & c_\theta c_{13}c_{23} & s_\theta \\
    s_\theta\left(c_{12}c_{23}s_{13}-s_{12}s_{23}\right) & s_\theta\left(c_{23}s_{12}s_{13}+c_{12}s_{23}\right) & -s_\theta c_{13}c_{23} & c_\theta
  \end{array}\right),
\end{aligned}
\end{equation}
with $c_\theta\equiv \cos\theta_\tau$, $s_\theta\equiv \sin\theta_\tau$ and $c_{ij}\equiv \cos\theta_{ij}$, $s_{ij}\equiv \sin\theta_{ij}$. For simplicity, we have ignored possible $CP$-violating phases in the mixing matrix.

Although we have four flavors of neutrino, our analysis of existing constraints differs from existing sterile neutrino analyses because the fourth mass eigenstate is assumed to be heavier than several MeV, in order to avoid cosmological constraints from the CMB~\cite{Ade:2013zuv} and BBN~\cite{Iocco:2008va}.
As discussed above, the heavy fourth mass eigenstate is mostly comprised of sterile and tau flavors, in order to satisfy laboratory and precision electroweak constraints on electron and muon neutrino mixing with a neutral heavy lepton. Current  terrestrial experiments produce either $\mu$ or $e$ flavor neutrinos at the source, and so can only  produce a linear combination of the  three light mass eigenstates. Since the light mass eigenstates are comprised of all four flavors, $e$, $\mu$, $\tau,$ and sterile $N_2$, the presence of the sterile component could affect neutrino oscillation experiments. Oscillations via the heavy neutrino are independent of the particular value of its mass since they correspond to a length
\begin{equation}
L=\frac{4\pi p}{\Delta m^2}\simeq \frac{4\pi p}{m_4^2}\lesssim 2.5\times 10^{-12}~{\rm cm}\left(\frac{p}{\rm MeV}\right),
\end{equation}
where $p$ is the momentum of the neutrinos in question, using the lower bound on the heavy neutrino mass of about 10~MeV. Therefore, the light mass differences must be given by the solar and atmospheric mass splittings as usual~\cite{Gonzalez-Garcia:2014bfa},
\begin{align}
\Delta m_{12}^2=\Delta m_\odot^2\simeq 7.5\times 10^{-5}~{\rm eV}^2,~\left|\Delta m_{13}^2\right|=\Delta m_{\rm atm}^2\simeq 2.5\times 10^{-3}~{\rm eV}^2,
\end{align}
where $\Delta m_{ij}^2\equiv m_i^2-m_j^2$.

We begin by noting that our assumption that $U_{e4}$ and $U_{\mu 4}$ are negligible allows us to determine $\theta_{12}$, $\theta_{13}$, and $\theta_{23}$
using terrestrial neutrino experiments which are insensitive to $\theta_\tau$. 
Combining these measurements with solar neutrino experiments allows for possible sensitivity to $\theta_\tau$. We describe this procedure below.

In principle, one needs to account for the effect of the different interaction with matter of the sterile neutrinos~\cite{Wolfenstein:1977ue,Mikheev:1986gs}, which can be included via a potential $V_{\rm nc}$ in the flavor basis for the active neutrino flavors of
\begin{equation}
V_{\rm nc}=-\frac{G_F}{\sqrt{2}} n_n
\end{equation}
where $n_n$ is the neutron density, with the opposite sign for antineutrinos. In matter with equal numbers of protons and neutrons and density of 2.7 $\rm g/cm^3$, a length scale of 4000 km can be derived from $1/V_{\rm nc}$,  which gives an estimate of the distance  scale required for   matter interactions to have an important effect in the analysis of neutrino oscillations~\cite{Pontecorvo:1967fh}  in the Earth's crust. Currently, the strongest constraints on the active neutrino mixing parameters derive from experiments which are not at long enough baseline to be highly sensitive to the matter effects. However, as we point out below, a strong limit on $\theta_\tau$ may be extracted from the IceCube and Super-Kamionkande experiments, due the difference in matter effects between sterile and active neutrinos as they travel through the Earth.

The best determination of $U_{e3}$ is by the reactor experiment Daya Bay~\cite{An:2013zwz}, which measures electron antineutrino  disappearance over a distance of 1.6 km. The value of $U_{\mu3}$ may be determined by measurements of muon neutrino and antineutrino disappearance by the K2K~\cite{Wilkes:2002mb} and MINOS experiments~\cite{Adamson:2011ig}, with baselines of 250 km and  730 km, respectively. Because matter effects are not highly significant at these baselines, extraction of this parameter is little affected by the possible presence of a sterile component in the third mass eigenstate. $U_{e2}$ can be determined by the long baseline reactor experiment KamLAND~\cite{Eguchi:2002dm} which has a baseline average of 180 km. These measurements of $U_{e3}$, $U_{\mu3}$, and $U_{e2}$ can be combined to give determinations of $\theta_{12}$, $\theta_{13}$, and $\theta_{23}$ that are independent of $\theta_\tau$, and these angles must be close to the values given by the usual three neutrino fits to data~\cite{Gonzalez-Garcia:2014bfa},
\begin{align}
\theta_{12}\sim 32^\circ,~\theta_{13}\sim 8^\circ,~\theta_{23}\sim 40^\circ.
\end{align}

Turning now to solar neutrinos, we note that  electron neutrino disappearance is mainly governed by the flavor composition of the second mass eigenstate, since  we may neglect the small angle $\theta_{13}$. High energy solar electron neutrinos are produced in the core of the sun, primarily via   $^8$B decays, and have a large effective mass from the matter interactions, larger than $\sqrt{\Delta m_{12}^2}$ but smaller than $\sqrt{|\Delta m_{23}^2|}$. These electron neutrinos are approximately an energy eigenstate of the effective Hamiltonian which includes  matter interactions. Adiabatic evolution of the electron neutrinos as they exit the core causes them to exit the sun primarily as the second mass eigenstate in vacuum. Neglecting $\theta_{13}$, the fraction of this mass eigenstate which is detected as electron neutrino is  $|s_{12}|^2$, while the fraction $|c_{12}s_{23}\sin\theta_\tau|^2$ is undetectable sterile. Hence, high energy charged current electron neutrino detection from solar neutrino experiments and the  solar neutral current flux~\cite{Aharmim:2011vm} can be used to constrain $\theta_\tau$ when combined with either the KamLAND determination of $\theta_{12}$ or with the theoretical calculation of the $^8$B flux. Because the KamLAND experiment, while very constraining of $\sqrt{\Delta m_{12}^2}$, is not as sensitive to $U_{e2}$, the theoretical calculation of the $^8$B flux currently gives the best precision on this determination. Experimentally, a combination of electron scattering and neutral current measurements are used  to calibrate the flux~\cite{Bellini:2008mr,D'Angelo:2014vgk} and the sterile neutrino component could affect this. Since the $^8$B flux is theoretically known to about the $15\%$ level currently~\cite{Serenelli:2011py}, we obtain a limit of $\left|\sin\theta_\tau\right|\lesssim 0.6$. This agrees with analyses of the combined solar data and KamLAND that has shown that the probability of electron neutrino disappearance into sterile neutrinos could be substantial~\cite{Dev:2005px,Cirelli:2004cz}.

Strong limits on $\theta_\tau$ come from the change in matter effects due to mixing with the sterile neutrino. As mentioned above, the light eigenstates are made up of $\nu_e$, $\nu_\mu$, and $\nu_{\tau N}$. In the presence of nonzero $\theta_\tau$, $\nu_{\tau N}$ has diminished weak interactions compared to $\nu_\mu$, with a potential given by $V_{\tau N}=V_{\rm nc}\cos^2\theta_\tau$. A recent search by Super-Kamiokande~\cite{Abe:2014gda} used atmospheric neutrinos to look for $\mu\to$~sterile transitions. Because of the lack of matter effects for the sterile neutrinos, this can manifest as a change in the distribution of muon neutrino zenith angle in the detector from the standard $\mu\to\tau$ transition scenario. At 90\%~C.\,L., the limit is $\left|U_{\tau4}\right|=\left|\sin\theta_\tau\right|< 0.42$. Similar effects were searched for in data from IceCube and DeepCore, using the language of neutrino nonstandard interactions (NSI)~\cite{Esmaili:2013fva}. In NSI studies, the difference in weak interaction strength between the light flavor eigenstate $\nu_{\tau N}$ and that of $\nu_\tau$ is parameterized by $\epsilon_{\tau\tau}$ with 
\begin{equation}
\epsilon_{\tau\tau}=\frac16\left(\frac{V_{\tau N}}{V_{\rm nc}}-1\right)=\frac{\sin^2\theta_\tau}{6}.
\end{equation}
In \REF~\cite{Esmaili:2013fva}, a 90\%~C.\,L. of $\epsilon_{\tau\tau}<0.03$ from azimuthal distributions of neutrinos was found. This also translates into $\left|U_{\tau4}\right|< 0.42$. We show this limit along with those from $\tau$ decays in \FIG~\ref{fig:Ulimit}.


\subsection{New couplings to the $Z$ and Higgs}
\label{sec:Zh}
\begin{figure}[tbp]
\centering
\includegraphics[width=0.8\textwidth]{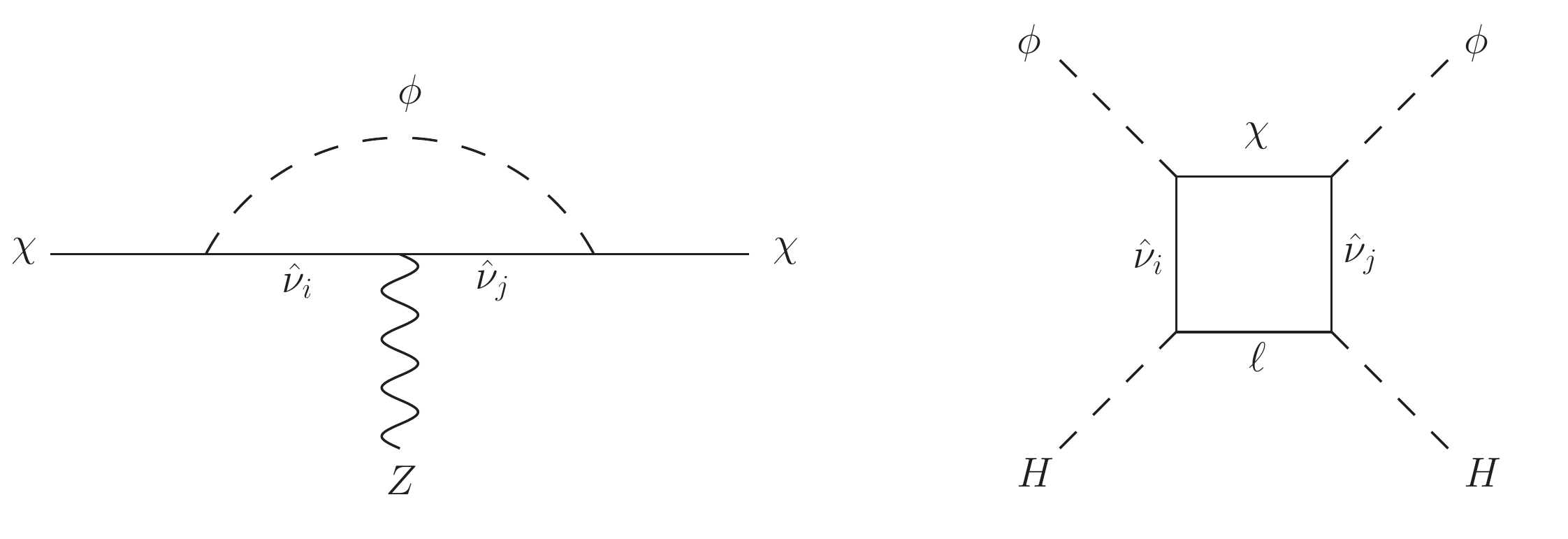}
\caption{\label{fig:Zh} Left: one loop diagram which gives rise to an effective $\chi$-$\bar\chi$-$Z$ coupling. Right: diagram that contributes to the $\left|\phi\right|^2\left|H\right|^2$ operator.}
\end{figure}

At one loop, through the diagram shown on the left in \FIG~\ref{fig:Zh}, an effective coupling of DM to the $Z$ boson is generated.
In the limit that $m_4\gg m_\phi$, this effective interaction is
\begin{align}
{\cal L}_{Z\bar\chi\chi}=-\frac{g_{\rm w}}{\cos\theta_{\rm w}}\left(\frac{y_2\sin2\theta_\tau}{8\pi}\right)^2Z_\mu\bar\chi_L\gamma^\mu\chi_L,
\end{align}
where $g_{\rm w}$ is the weak coupling strength and $\theta_{\rm w}$ is the weak mixing angle. This operator contributes to the invisible decay width of the $Z$ (the rate for $Z\to$~neutrinos is unchanged due to the unitarity of $U$ for $m_4\ll m_Z/2$). The good agreement between the SM expectation for this rate and experiment offers a potential constraint on the model. The rate for $Z\to\chi\bar\chi$ through this operator is
\begin{align}
\Gamma_{Z\to\chi\bar\chi}=4.2\times 10^{-4}\left(y_2\sin2\theta_\tau\right)^4~{\rm MeV}.
\end{align}
The 95\%~C.\,L. upper limit on extra contributions to the $Z$ invisible width is $2.0~\rm MeV$~\cite{ALEPH:2005ab}. This translates to a weak limit of $\left|y_2\sin2\theta_\tau\right|<8.3$.

This effective $Z_\mu\bar\chi_L\gamma^\mu\chi_L$ interaction can lead to the scattering of DM on normal matter. For phenomenologically interesting values of the parameters, however, this scattering is highly suppressed: taking $y_2=1$ and $\theta_\tau=0.3$, the cross section is about $10^{-7}-10^{-6}$ times that of neutrino scattering. Because this cross section is so small, using proton beam dumps to produce the heavy neutrino, though production of $D_s\to\tau\nu_\tau$ (as used in searches for visibly decaying heavy neutrinos~\cite{Orloff:2002de,Astier:2001ck,Abreu:1996pa}), which decays to DM that scatters in a detector (see, e.g.~\cite{Batell:2009di,deNiverville:2011it,deNiverville:2012ij,Batell:2014yra}) is not promising.

We now consider interactions involving the Higgs. The operators proportional to $\lambda_i$ in \eqref{eq:Lm} will lead to contributions to the invisible width of the Higgs boson after electroweak symmetry breaking, through $h\to N_1 \nu_i$. Since we ignore $\lambda_{e,\mu}$ and take $m_4\ll m_h=125~{\rm GeV}$, the rate for this decay is
\begin{align}
\Gamma_{h\to{\rm inv.}}=\frac{\lambda_\tau^2}{16\pi}m_h.
\end{align}
The invisible branching ratio of the Higgs is presently limited to about 25\%~\cite{Giardino:2013bma}, which translates into $|\lambda_\tau|\lesssim 2\times10^{-2}$. This is not constraining on the model since $|\lambda_\tau|=(m_4/v)|\sin\theta_\tau|$ is less than $2\times10^{-3}$ if $m_4<300~\rm MeV$.

At one loop, as seen on the right in \FIG~\ref{fig:Zh}, a logarithmically-divergent dimension-4 operator involving the scalar $\phi$ and the Higgs doublet is generated,
\begin{align}
{\cal L}_{\phi H}&=\lambda_{\phi H}\left|\phi\right|^2\left|H\right|^2,
\end{align}
with
\begin{align}
\lambda_{\phi H}&\sim \left(\frac{y_1\lambda_\tau}{2\pi}\right)^2\log\left(\frac{\Lambda^2}{M^2}\right)=\frac{g^2}{4\pi^2}\left(\frac{y_1}{y_2}\right)^2\left(\frac{m_4}{v}\right)^2\log\left(\frac{\Lambda^2}{M^2}\right),
\end{align}
where $\Lambda$ is the scale of the physics that enters to cut this contribution off. After electroweak symmetry breaking, this gives a contribution to the mass of $\phi$ given by $\delta m_\phi^2=\lambda_{\phi H}v^2$. This contribution defines a lower bound on $m_\phi$; obtaining a mass below this value requires some fine-tuning of this contribution against the bare value of the mass. Noting that $M\simeq m_4$ and choosing $\Lambda=1~\rm TeV$,
\begin{align}
\delta m_\phi^2&\sim \left(10~{\rm MeV}\right)^2\left(\frac{g}{0.3}\right)^2\left(\frac{y_1/y_2}{0.5}\right)^2\left(\frac{m_4}{100~\rm MeV}\right)^2\left[\log\left(\frac{1~{\rm TeV}}{m_4}\right)\Big/ 10\right].
\end{align}
Therefore, $\phi$ can have a mass in the tens of $\rm MeV$ range for a heavy neutrino with mass of ${\cal O}(100~{\rm MeV})$ without running into any fine-tuning problems, even for a cutoff at the $\rm TeV$ scale.


\section{Solving the missing satellites problem}
\label{sec:sss}
As was mentioned in the introduction, DM-neutrino interactions will suppress the growth of small scale DM density perturbations in the early Universe, helping to alleviate the missing satellites problem.  This suppression of small scale structure occurs below the maximum of two different length scales for washing out structure.  These two length scales have different physical origins and will be discussed in detail in this section.  Since the DM density is known, this maximum length scale corresponds to a mass cutoff scale, $M_{\rm cutoff}$, below which the formation of less massive structures is suppressed.

The first scale for washing out small scale structure is set at early times when the DM is in thermal equilibrium with the relativistic plasma.  Once $T \lesssim m_{\chi}$ the expansion of the Universe will cause the plasma density to decrease enough such that the annihilation and production scattering processes keeping DM in chemical equilibrium with the plasma will freeze out, ending DM number-changing processes. (Note that DM-neutrino interactions that are strong enough to solve the missing satellites problem force DM to be asymmetric and not a thermal relic--see the discussion following \EQ~\eqref{eq:ann_xsec}.)  However, DM-neutrino elastic scattering can keep the DM in thermal equilibrium even after chemical decoupling.  The DM eventually will fall out of thermal equilibrium once the DM-neutrino elastic scattering rate drops below the Hubble expansion rate of the Universe.  This time when elastic scattering ceases is called kinetic decoupling.  After this, the DM simply free-streams and washes out small scale structure, setting another scale below which structure formation is suppressed.   

In the following we discuss how DM-neutrino interactions can lead to a value of $M_{\rm cutoff}$ in the range of $10^7 M_{\odot} - 10^9 M_{\odot}$ which is large enough to solve the missing satellites problem.   We will find the range of interesting DM-neutrino coupling, $g$, and DM and mediator masses, $m_{\chi}$ and $m_{\phi}$, which can achieve these cutoff mass scales.

\subsection{Kinetic decoupling condition}
Kinetic decoupling occurs when the rate for DM-neutrino collisions to change the DM momentum, $\gamma(T)$, becomes small compared to the Hubble parameter, $H(T)$.
Hence the decoupling temperature, $T_{\rm d}$, can be estimated by solving
\begin{equation}
\label{eq:kd}
\gamma(T_d)  = H(T_d),
\end{equation}
where
\begin{equation}
\gamma(T) = \frac{1}{3m_{\chi}T}\int_0^{\infty}\frac{d^3p}{(2\pi)^3}f(p/T)(1-f(p/T))\int_{-4p^2}^0dt(-t)\frac{d\sigma_{\nu \chi}}{dt}.
\end{equation}
Here $f(p/T) = (e^{p/T}+1)^{-1}$ is the Fermi-Dirac distribution function describing the neutrinos in the massless limit and $t$ is the usual Mandelstam variable.  This kinetic decoupling equation comes from an approximate solution to the Boltzmann equation, see~\cite{Gondolo:2012vh}.  As we will soon show, DM and the neutrinos must remain in kinetic equilibrium until $T \simeq 1$ keV, which occurs after the neutrinos decouple from the photons at $T \simeq 1$ MeV, in order for DM to solve the missing satellites problem. This means that the terms on the LHS of \EQ~\eqref{eq:kd} which have to do with the neutrino-DM fluid should be evaluated at the neutrino temperature, which differs from the photon temperature via $T_{\nu} = (4/11)^{1/3}T_{\gamma}$.  In what follows all temperatures are photon temperatures and the factor of $(4/11)^{1/3}$ has been included when necessary.

We solve for $T_{\rm d}$ using the approximate form of the total DM-neutrino cross section for $E_{\nu} \ll m_{\chi},~m_{\phi}$.  This is a good approximation near decoupling since at this point $E_{\nu} \sim T \ll m_{\chi},~m_{\phi}$.  Hence
\begin{equation}
\frac{d \sigma_{\nu \chi}}{dt} = \frac{g^4}{32 \pi \left(m_{\phi}^2-m_{\chi}^2\right)^2}.
\end{equation}
Using this in \EQ~\eqref{eq:kd}, the remaining integrals can be done analytically and the decoupling temperature is given by
\begin{equation}
\begin{aligned}\label{eq:Td}
T_{\rm d} &= \left(\frac{5082}{31\pi\sqrt{5\pi}}\right)^{1/4}\left(\frac{g_{\rm eff}(T_d)^{1/8}}{M_{\rm Pl}^{1/4}}\right)\left(\frac{m_{\chi}^{1/4}\sqrt{m_{\phi}^2-m_{\chi}^2}}{g}\right) \\
&= 1.6 ~\text{keV}~\left(\frac{g_{\rm eff}(T_{\rm d})}{3.36}\right)^{1/8} \left(\frac{m_{\chi}}{20 \text{ MeV}}\right)^{1/4}\left(\frac{\sqrt{m_{\phi}^2-m_{\chi}^2}}{35 \text{ MeV}}\right)\left(\frac{g}{0.3}\right)^{-1}, 
\end{aligned}
\end{equation}
where $g_{\rm eff}(T)$ is the effective number of relativistic, bosonic degrees of freedom at temperature $T$.  This expression for the decoupling temperature contains the correct parametric dependence derived using simple arguments in the introduction.  Note that we used the expression for the Hubble parameter during the radiation dominated period (valid down to $T \simeq 1$ eV) given by
\begin{equation}
H = \sqrt{\frac{4\pi^3g_{\rm eff}(T)}{45 M_{\rm Pl}^2}}T^2.
\end{equation}

\subsection{The cutoff mass scale}
There are two main processes that erase primordial density fluctuations in the DM fluid on small scales: (i) acoustic oscillations in the coupled, relativistic plasma of the early universe up until the time of kinetic decoupling, and (ii) free streaming of DM after kinetic decoupling.  The larger of the two scales set by these processes determines $M_{\rm cutoff}$.

While DM remains in thermal equilibrium with the relativistic plasma, it is involved in the acoustic oscillations of the plasma since it couples to the neutrinos.  This results in damped oscillations in the DM power spectrum that appear on the scale of the horizon at kinetic decoupling, $H_{\rm d}^{-1} = a_{\rm d}\eta_{\rm d}$, where $\eta_{\rm d} = \int_0^{t_{\rm d}} dt/a(t)$ is the comoving distance a photon can travel from the beginning of the Universe until the time of kinetic decoupling~\cite{Loeb:2005pm}.  Here $a(t)$ denotes the scale factor in a Friedmann-Robertson-Walker metric and $a_{\rm d}$ is the scale factor at the time of kinetic decoupling.  This smallest distance scale corresponds to a DM halo mass cutoff given by
\begin{equation}
M_{\rm ao} = \rho_{\chi}(T_{\rm d})\frac{4\pi}{3} (a_{\rm d}\eta_{\rm d})^3 ,
\end{equation}
where $\rho_{\chi} (T)$ is the DM energy density (equal to its mass density for $T < T_{\rm d} \ll m_{\chi}$) at temperature $T$.  Since the mass enclosed in a given volume remains the same even as that volume expands, $M_{\rm ao}$ can also be expressed in terms of the DM density and scale factor today as
\begin{equation}
M_{\rm ao} = \rho_{\chi}(T_0)\frac{4\pi}{3} (a_0\eta_{\rm d})^3 .
\end{equation}
Using typical values and assuming entropy in a comoving volume is conserved from $T_{\rm d}$ until today, this becomes
\begin{equation}\label{eq:Mao}
M_{\rm ao} = 2 \times 10^{8}M_{\odot} ~\left(\frac{g_{\rm eff}(T_{\rm d})}{3.36}\right)^{-1/2}\left(\frac{T_{\rm d}}{\text{\text{keV}}}\right)^{-3},
\end{equation}
where we used $H_0 = 67$ km/s/Mpc, $\Omega_{\chi} = 0.27$, $g_{\rm eff}(T_0) = 3.36$ and $T_0 = 2.7$ K.

After kinetic decoupling, DM free-streams, washing out structure on scales smaller than $\ell_{\rm eq} = \pi a_{\rm eq}\int_{t_{\rm d}}^{t_{\rm eq}}dt(v_{\rm phys}/a(t))$ at the time of matter-radiation equality~\cite{Loeb:2005pm, Green:2005fa}.  Here $a_{\rm eq}$ is the scale factor at matter-radiation equality, $v_{\rm phys} = v/a(t)$ is the velocity of the DM particles, and $v$ is their constant comoving velocity.  This scale describes the distance that DM free-streams from $T_{\rm d}$ to $T_{\rm eq} \simeq 1$ eV.  Up until $T_{\rm eq}$ this scale grows as $\ln{T}$ and the growth after $T_{\rm eq}$, proportional to $T^{-1/3}$, has been neglected.  Evaluating $\ell$ today we find
\begin{equation}
\ell_0 = \left(\frac{a_0}{a_{\rm d}}\right)\left(\frac{v}{a_{\rm d}}\right)\frac{\pi}{H_{\rm d}}\ln\left[\frac{g_{\rm eff}(T_{\rm d})^{1/3}T_{\rm d}}{g_{\rm eff}(T_{\rm eq})^{1/3}T_{\rm eq}}\right].
\end{equation}
Approximating the DM velocity at the time of decoupling as $v/a_{\rm d} = \sqrt{(4/11)^{1/3}T_{\rm d}/m_{\chi}}$, the cutoff mass scale due to DM free-streaming is given by
\begin{equation}
\begin{aligned}\label{eq:Mfs}
M_{\rm fs} =& ~\rho_{\chi}(T_0)\frac{4\pi}{3} \ell_0^3 \\
 =&~ 3 \times 10^5 M_{\odot} ~\left(\frac{g_{\rm eff}(T_{\rm d})}{3.36}\right)^{-1/2}\left(\frac{m_{\chi}}{20 \text{ MeV}}\right)^{-3/2}\left(\frac{T_{\rm d}}{\text{keV}} \right)^{-3/2} \\ 
&\hspace{30mm}\times\left\{1+\ln\left[\left(\frac{g_{\rm eff}(T_{\rm d})}{3.36}\right)\left(\frac{T_{\rm d}}{\text{keV}}\right)\right]/6.0\right\}^3.
\end{aligned}
\end{equation}

\begin{figure}[tbp]
\centering
\includegraphics[width=.45\textwidth]{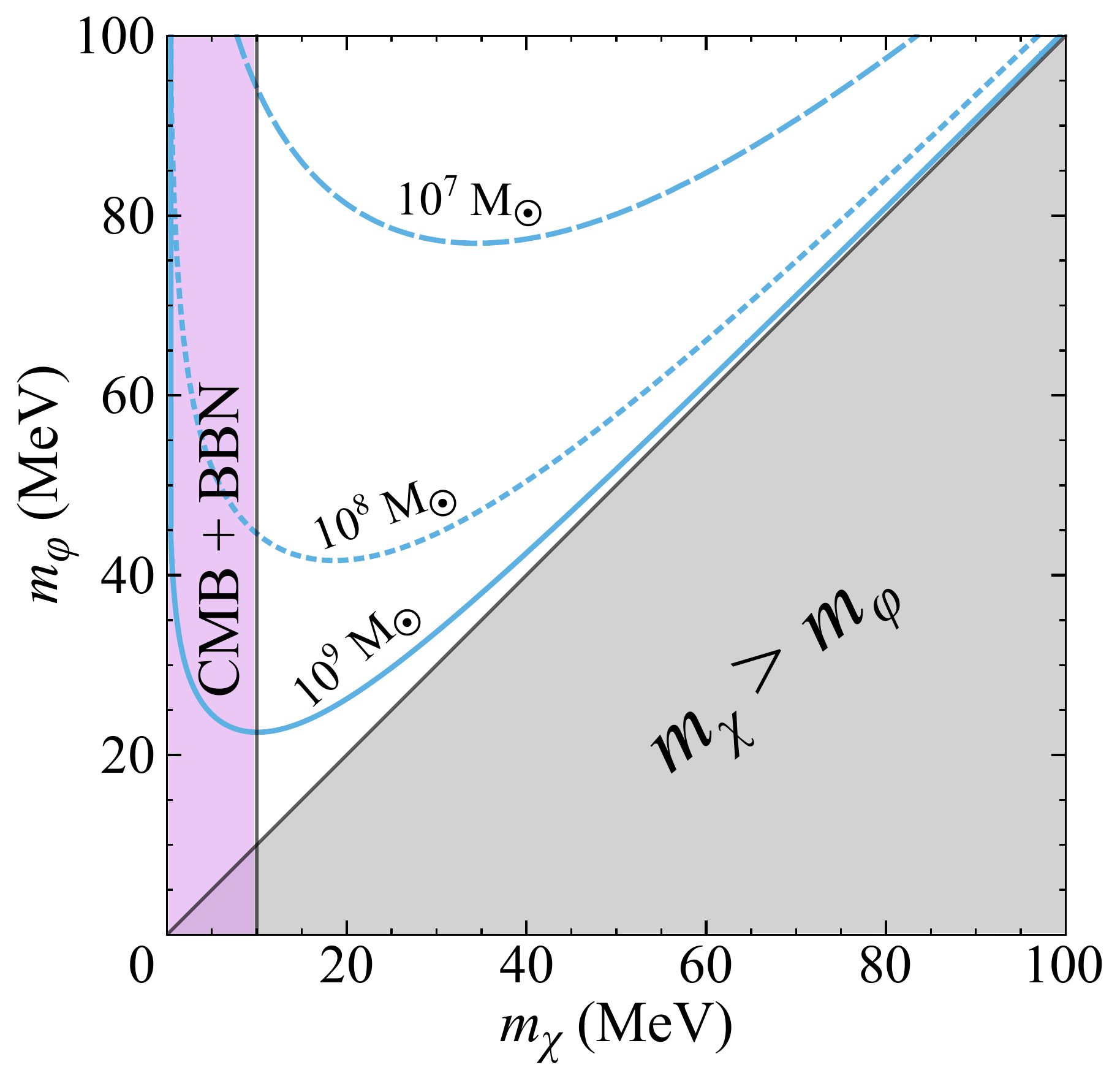}
\hfill
\includegraphics[width=.44\textwidth]{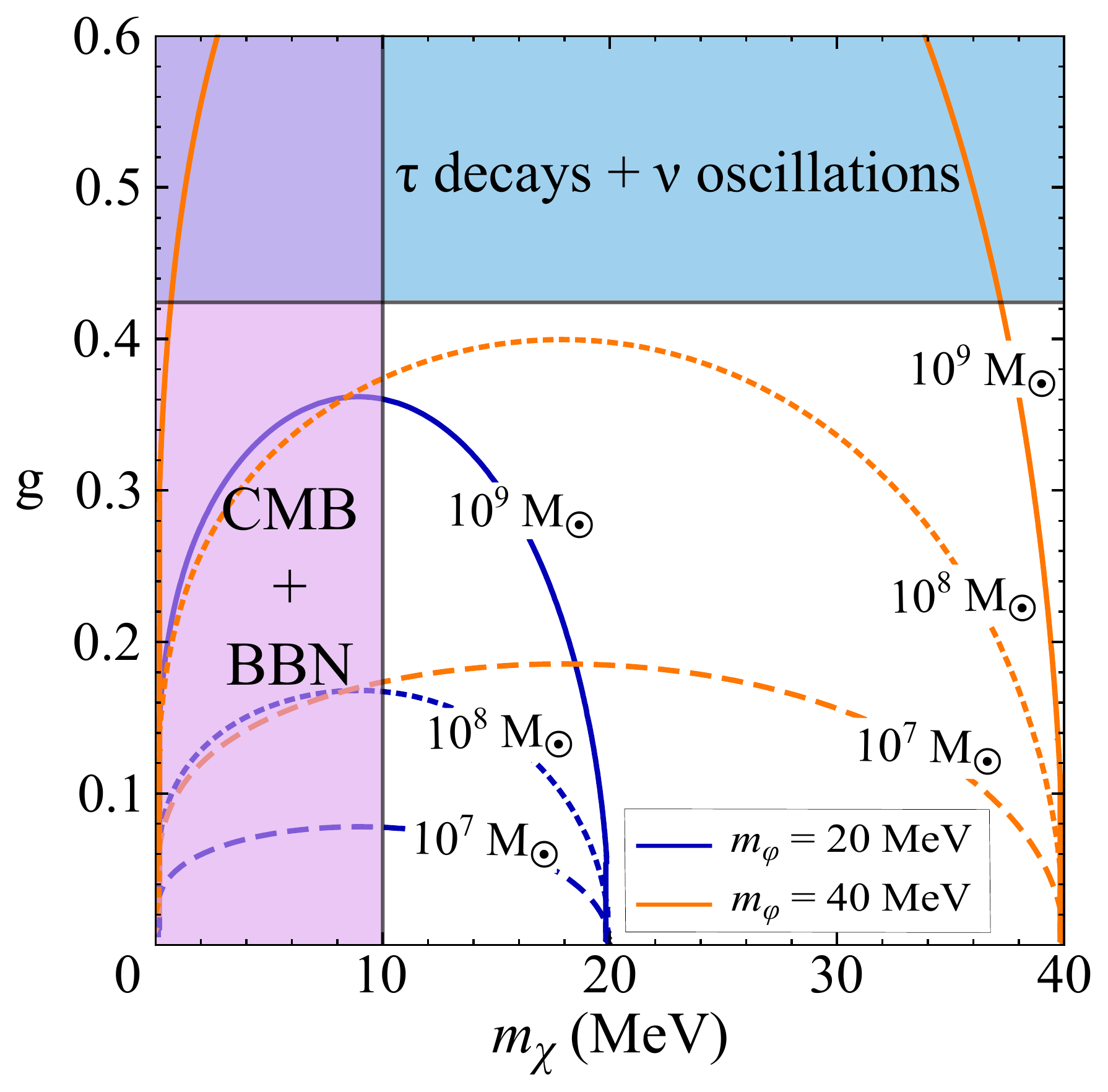}
\caption{\label{fig:mphivsmchi} Left: Values of $m_\phi$ required for $M_{\rm cutoff}=10^7$ (dashed), $10^8$ (dotted), and $10^9M_\odot$ (solid) as functions of $m_\chi$. To fix $g=y_2\left|U_{\tau4}\right|$, we set $y_2=1$ and take the largest value of $\left|U_{\tau4}\right|$ allowed by $\tau$ decays and neutrino oscillation experiments, $0.42$, as shown in \FIG~\ref{fig:Ulimit}. The gray shaded region on the bottom-right corresponds to the unphysical situation where the mediator is lighter than the DM. Right: the coupling $g$ required for $M_{\rm cutoff}=10^7$, $10^8$, and $10^9M_\odot$ varying the DM mass for $m_\phi=20$ and $40~\rm MeV$. The upper limit on $g=y_2\left|U_{\tau4}\right|$ of 0.42 (c.f. \SEC~\ref{sec:Ulimits}) assuming $y_2=1$ is also shown. In both plots we show the lower limit on the DM mass from observations of the CMB and BBN.}
\end{figure}

The smallest mass object formed by DM is the largest of $M_{\rm ao}$ and $M_{\rm fs}$.  Comparing \EQ~\eqref{eq:Mao} and \EQ~\eqref{eq:Mfs}, we see that in order to obtain values of $M_{\rm cutoff}$ in the range $10^7 - 10^9 M_{\odot}$ with $m_{\chi,\phi} \sim \text{few} \times 10$ MeV, $T_{\rm d} \sim$ keV is needed and acoustic oscillations set the cutoff scale.  Hence, combining \EQ~\eqref{eq:Td} and \EQ~\eqref{eq:Mao}, we have that
\begin{equation}
M_{\rm cutoff} = 4 \times 10^7 ~M_{\odot}~\left(\frac{g_{\rm eff}(T_{\rm d})}{3.36}\right)^{-7/8}\left(\frac{g}{0.3}\right)^3 \left(\frac{m_{\chi}}{20\text{ MeV}}\right)^{-3/4}\left(\frac{\sqrt{m_{\phi}^2-m_{\chi}^2}}{35 \text{ MeV}}\right)^{-3}.
\end{equation}
The left panel of \FIG~\ref{fig:mphivsmchi} shows the cutoff scale varying $m_\chi$ and $m_\phi$. We take $g=0.42$ to be as large as allowed by limits on $\left|U_{\tau4}\right|$ from $\tau$ decays and neutrino oscillation experiments with $y_2=1$. On the right panel we display the coupling $g$ required to obtain $M_{\rm cutoff}=10^7$, $10^8$, and $10^9M_\odot$ as a function of $m_\chi$ for $m_\phi=20$ and $40~\rm MeV$. We set $y_2=1$ and show the resulting upper limit on $g=y_2\left|U_{\tau4}\right|$ from the limit $\left|U_{\tau4}\right|<0.42$ as found in \SEC~\ref{sec:Ulimits}.

Finally, we note that the effect of DM-photon interactions on the nonlinear structure formation of satellite galaxies in a Milky Way sized galaxy has been simulated in~\cite{Boehm:2014vja}.  The effects of DM-photon interactions on structure formation should be very similar to the effects of DM-neutrino interactions since they both suppress structure formation on small scale due to acoustic oscillations. They find that for a constant DM-neutrino cross section, $\sigma_{\text{DM}-\gamma} \gtrsim 7 \times 10^{-35}$ cm$^2$ for $m_{\chi} = 20$ MeV, is ruled out at the $2$ sigma level since then DM-photon interactions would wash out too much structure to be consistent with the number of satellite galaxies that we observe in the Milky Way.  In our scenario, the DM-neutrino cross section is not a constant, but at the time of kinetic decoupling when the value of the DM-neutrino cross section is most important for affecting small scale structure, we have that for typical parameters, $\sigma \simeq g^4 (3 T_{\rm d})^2 / (8\pi (m_{\phi}^2-m_{\chi}^2)^2) \simeq 2 \times 10^{-36}$ cm$^{2}$, which is within the bounds of \REF~\cite{Boehm:2014vja}, but still large enough to significantly decrease structure formation on small scales. Similarly, in simulations of a model in which dark matter interacts with dark radiation \cite{Buckley:2014hja}, small galaxies form later and have lower central densities than in standard CDM.

\section{Implications for supernovae}
\label{sec:sn}
Supernovae (SNe), being abundant sources of neutrinos, can offer interesting information about strong DM-neutrino interactions. In this section we examine the effects of such interactions on the properties of SNe.

\subsection{Neutrino emission and cooling}
In the standard picture of core collapse SNe, the three flavors of neutrinos and antineutrinos are produced in the supernova (SN) at temperatures around 30 MeV, mainly via nucleon bremsstrahlung and electron neutrino-antineutrino annihilation.  Outside of the first neutronization burst of electron neutrinos, the neutrinos remain trapped in the dense core of the collapsing star for $\sim 0.2$ s at which point they free-stream out of the star over a time period of $\sim 10$ s, carrying away the binding energy of the remaining proto-neutron star $\sim 3 \times 10^{53}$ erg.  For recent reviews see e.g.,~\cite{Janka:2006fh,Janka:2012wk}.

DM candidates with $m_{\chi} \lesssim 100$ MeV are light enough to be thermally produced in SNe.  If these DM candidates are weakly interacting, then they can be constrained since the presence of this DM could help cool the proto-neutron star, producing a neutrino signal that is in conflict with the observations from SN 1987A.  In our scenario, DM (with a mass $\gtrsim 10$ MeV) will also be produced in the SN.  However, due to the strong DM-neutrino interactions, this DM will thermalize with the neutrino gas and maintain a thermal distribution out to large radii until the temperature of the neutrinos falls below the DM mass, suppressing DM production.

Neutrinos begin to free-stream away from the SN when the density of the stellar material drops, which occurs where the matter temperature is $\sim 5$ MeV $< m_{\chi}$.  At this point, DM production is suppressed, but the coupled DM-neutrino gas from the core will still diffuse out of the star, cooling the star on timescales set by the speed of sound in the DM-neutrino fluid.  In this sense, strong DM-neutrino interactions are similar to strong neutrino self-interactions in SNe since they both involve the emission of a strongly-coupled gas, and hence strong DM-neutrino interactions do not significantly affect the cooling time for SNe~\cite{Dicus:1988jh,Boehm:2013jpa}.  Thus it is likely that this neutrino-interacting DM will not come into conflict with the observation of neutrino cooling from SN 1987A.

In~\cite{Fayet:2006sa} a similar scenario of relatively strong DM-nucleon and DM-neutrino interactions inside SNe was considered.  In this case, the DM thermalized with the stellar material and bound the neutrinos to the star out to larger radii and lower temperatures (a result similar to what would be expected from simply increasing the strength of neutrino interactions with regular matter).  This would lead to an overall decrease in the energies of the emitted neutrinos and an increase in the cooling time, resulting in a rough bound of $m_{\chi} \gtrsim 10$ MeV in order to be consistent with the neutrino observations from SN 1987A.  However, for the case of DM that only interacts with neutrinos, it is natural to expect that the constraint on $m_{\chi}$ will be weakened since the DM does not have strong interactions with the stellar material and is not trapped in the core of the star.  A precise study of the emission of neutrinos is beyond the scope of this work and will be explored in a future paper~\cite{future}.

Finally, \REFS~\cite{Boehm:2013jpa,Mangano:2006mp} find that the constraints on neutrino-interacting DM from SNe come not from cooling, but from SN neutrinos scattering off the DM and out of the line of sight of our detectors.  They place a bound on the DM-neutrino cross section of $\sigma_{\hat{\nu}_i \chi} \lesssim 10^{-25}$ cm$^2$ $\left(m_{\chi}/\text{MeV}\right)$ by requiring that the neutrino mean free path be larger than the Earth-SN distance for a nearby SN.  In the next section, we will find that our neutrino-DM cross section abides by this constraint except near resonance, producing a feature in the neutrino spectra which should be observable in the next galactic SN. 

\subsection{Observation of a nearby supernova}
\label{sec:obsSN}
An interesting consequence of strong DM-neutrino interactions is the scattering of SN neutrinos off DM on their way to Earth. The parameters implied by the missing satellite problem make this particularly intriguing because the resonant neutrino energy for scattering, $E_{\rm res}=(m_\phi^2-m_\chi^2)/2m_\chi$, is in the range of energies produced in SNe since both $\chi$ and $\phi$ have masses that are tens of MeV.

We consider a light neutrino mass eigenstate $i$ that was emitted from a SN. As it travels from the SN to Earth, scattering on DM can deflect it, decreasing the flux that is observed,
\begin{align}
{\rm Flux}\left(\hat\nu_i\right)_{\rm Earth}={\rm Flux}\left(\hat\nu_i\right)_{\rm SN}e^{-\Gamma_i d},
\end{align}
where
\begin{align}
\Gamma_i=\sigma_{\hat\nu_i\chi}\times\frac{1}{d}\int_0^d dx\, n_\chi.
\end{align}
As defined in \eqref{eq:sigmai}, $\sigma_{\hat\nu_i\chi}$ is the cross section for $\hat\nu_i$ to scatter on DM at rest, $d$ is the distance between the SN and Earth, and $n_\chi$ is the DM number density along the line of sight. Using \eqref{eq:sigmai}, we can isolate the mass eigenstate dependence,
\begin{align}
\Gamma_i=\frac{\left|U_{Ni}\right|^2}{\left|U_{e4}\right|^2+\left|U_{\mu 4}\right|^2+\left|U_{\tau 4}\right|^2}\Gamma,
\end{align}
with 
\begin{align}
\Gamma=\sigma_{\nu\chi}\times\frac{1}{d}\int_0^d dx\, n_\chi\left(x\right).
\end{align}
To get a rough idea of what distance scale this attenuation occurs on, we set $n_\chi(x)$ to a constant value, with a magnitude equal to the local DM density, which is typical on galactic scales. That is, we take $n_\chi(x)=\bar n_\chi=(0.3~{\rm GeV}/m_\chi)~{\rm cm^{-3}}$. Then, $1/\Gamma=1/\sigma_{\nu\chi}\bar n_\chi$ defines a length scale over which the scattering of DM is important. 
This length scale can be comparable to galaxy sizes for neutrinos with energy close to the resonance energy, $E_{\rm res}$. At this energy the cross section is, for $m_\phi\gg m_\chi$,
\begin{align}
\sigma_{\nu\chi}\simeq\frac{4\pi}{m_\phi^2}=3\times10^{-24}~{\rm cm^2}\left(\frac{40~\rm MeV}{m_\phi}\right)^2,
\end{align}
where we have also assumed that $m_\phi<m_4$ so that $\Gamma_\phi=g^2m_\phi/16\pi$. This cross section leads to an attenuation length
\begin{align}
\frac{1}{\Gamma}\simeq 7~{\rm kpc}\left(\frac{m_\phi}{40~\rm MeV}\right)^2\left(\frac{m_\chi}{20~\rm MeV}\right).
\end{align}
The cross section is this large only in a region of width ${\cal O}(\rm MeV)$ around $E_{\rm res}$. Off resonance, the cross section drops quite rapidly below the bound found in~\cite{Boehm:2013jpa,Mangano:2006mp}. Therefore, the DM-neutrino interactions show up as a feature in the spectrum of neutrinos from a SN at an energy given by $E_{\rm res}=(m_\phi^2-m_\chi^2)/2m_\chi$.

The mixing matrix $U$ determines the relative attenuation of each eigenstate. For simplicity, in the tribimaximal approximation which is a good rough description of the neutrino mixing pattern, $\sin\theta_{12}=1/\sqrt3$, $\sin\theta_{23}=1/\sqrt3$, and $\theta_{13}=0$, the attenuation scales for the three light eigenstates are
\begin{align}
\frac{1}{\Gamma_1}\simeq\frac{6}{\Gamma},\quad\frac{1}{\Gamma_2}\simeq\frac{3}{\Gamma},\quad\frac{1}{\Gamma_3}\simeq\frac{2}{\Gamma}.
\end{align}
Because of this hierarchy, the fraction of $\hat\nu_1$ neutrinos is increased due to scattering on DM. As for the flavor composition of the neutrinos, because $\hat\nu_1$ has a larger component of $\nu_e$ than $\hat\nu_2$ or $\hat\nu_3$, the fraction of electron neutrinos detected from a SN is likewise increased. Thus, an increase in the electron neutrino fraction at $E_{\rm res}$ is a telltale sign of strong DM-neutrino interactions.

Beyond affecting the signals from nearby SNe, neutrino-DM interactions can leave an imprint in the diffuse SN background (DSNB). This was studied in detail in~\cite{Farzan:2014gza} in the context of an effective interaction between (scalar) DM and neutrinos. For parameters relevant for our scenario, the spectral distortion at $E_{\rm res}$ could be observable at proposed next generation experiments like Hyper-Kamiokande.\footnote{In a similar vein (although unconnected to SNe), neutrinos with energies of $10^2 - 10^3~{\rm TeV}$ have been explored as probes of new neutrino interactions due to scattering~\cite{Blum:2014ewa}. Because of the suppression of the scattering cross section at high energies, $\sigma_{\nu\chi}\propto1/E_\nu$, and the small DM density relevant for neutrinos traveling cosmological distances, $\rho_\chi\sim1.5~{\rm keV}/{\rm cm}^3$, this is unimportant in our scenario.}

\subsection{The "core vs. cusp" and "too big to fail problems"}
SNe can also figure prominently in potential solutions to the "core vs. cusp" problem.  This problem arises from the discrepancy in the DM density profile near the centers of galaxies between standard CDM simulations, which predict cusps, and observations, which favor cores. Some simulations~\cite{Governato:2009bg,Governato:2012fa,Teyssier:2012ie,Zolotov:2012xd,Madau:2014ija} and analytic models~\cite{Pontzen:2011ty,Penarrubia:2012bb} that include feedback from SNe on the DM indicate that such a coupling can modify the shape of DM profiles near the centers of galaxies, hence solving the core vs. cusp problem.  The energy transferred from SNe to the interstellar medium modifies the gravitational potential felt by the DM, allowing the DM to move away from the center of the galaxy, creating a more cored profile. In some simulations, taking reasonable values for the SN rate, transferring on the order of $10^{50}-10^{51}~\rm ergs$ per SN to the DM is sufficient to turn a cusped halo into a cored one~\cite{Governato:2009bg,Madau:2014ija}.

Additionally, SN feedback can address the "too big to fail problem," in which simulations predict that the Milky Way satellite galaxies should be more massive than they are observed to be.  SN feedback reduces the DM density in the center of galaxies, and this also helps solve the too big to fail problem \cite{2014arXiv1408.6444O}.  In~\cite{Madau:2014ija}, N-body simulations including SN feedback showed that indeed the too big to fail problem could be solved by SN feedback moving DM from the center region of galaxies out to larger radii.

It has also been suggested that SN feedback may not be sufficient to address these small scale structure problems~\cite{BoylanKolchin:2011dk,GarrisonKimmel:2013aq}. Maximally, around $1\%$ of the supernova energy can be transferred to DM gravitationally, causing the DM to move away from the center of galaxies, via the method described above.  The other $99\%$ of the SN energy is released in the form of neutrinos. In the case of strong DM-neutrino interactions, the neutrinos released by a SN can transfer energy to DM by elastic scattering.  This increases the transfer of energy from SNe to DM and potentially makes SN feedback more effective at solving the core vs. cusp and too big to fail problems.

Simple estimates show that the energy transfer from SNe to DM through this mechanism is of the right order of magnitude to solve the core vs. cusp problem. However, since the scattering length of the neutrinos is a kpc or larger, as seen in \SEC~\ref{sec:obsSN}, each neutrino emitted by a SN in the inner region of a galaxy scatters at most once an ${\cal O}(1)$ number of times as it leaves the galaxy. In, for example, a bright dwarf galaxy like Fornax, it is estimated that about $10^5$ SNe have occurred~\cite{GarrisonKimmel:2013aq}, each of which emitted about $10^{58}$ neutrinos so that maximally around $10^{63}$ DM particles will gain energy from SNe neutrinos through scattering. This should be compared to the roughly $10^{68}$ DM particles in Fornax, given a galactic mass of $10^9 M_{\odot}$ and a 20~MeV DM mass. Therefore, the energy from SNe is only distributed to a small fraction of the DM and cannot turn a core into a cusp. Accounting for neutrinos from stars that do not become SNe could have an effect on the core vs. cusp problem and will be studied in future work~\cite{future}.


\section{Future tests}
\label{sec:future}
As we have described, to achieve a cutoff on DM structures of $M_{\rm cutoff}\sim10^8M_\odot$ requires $\left|U_{\tau4}\right|\gtrsim0.1$. One promising test of strong DM-neutrino interactions is to improve the searches that are sensitive to $\left|U_{\tau4}\right|$. We discuss prospects for this improvement below.

\subsection{$\tau$ decays}
\label{sec:taudecays}
For $m_4> 100~{\rm MeV}$ the strongest constraint on $\left|U_{\tau4}\right|$ comes from $\tau$ decays, in particular our estimate using changes to $\Gamma_{\tau\to\mu\nu\bar\nu}$. However, the measurements of the branching ratio for $\tau\to\mu$ were not searches for heavy neutrinos and could be subject to systematic biases in acceptance estimates that assume a vanishing neutrino mass. The best dedicated experimental search for a heavy component to $\nu_\tau$ used LEP data, based on about $10^5$ $\tau^+\tau^-$ pairs, looking at hadronic three- and five-prong decays~\cite{Barate:1997zg}. We strongly suggest that new experimental searches be undertaken to search for a massive (greater than $10~\rm MeV$) neutrino component of $\nu_\tau$. The B-factories have each collected about $10^4$ times more $\tau$ pairs and Belle II will improve on that by an order of magnitude. Therefore the statistical errors in such a search could conceivably improve by $\sim100$. Using several decay channels is a good strategy since multi-prong hadronic final states are more sensitive to the reduced phase space available but leptonic decays are subject to less theoretical uncertainty. Although the search in \REF~\cite{Barate:1997zg} was systematics limited, if the systematic errors for new dedicated searches can be controlled to the level of the statistical ones, an improvement of the sensitivity to $\left|U_{\tau4}\right|$ by a factor of $10$ would be possible, exploring a large amount of parameter space favored by solutions to small scale structure problems. Improving these searches would be a highly desirable test of DM-neutrino interactions.

We also briefly mention here that the value of $\left|V_{us}\right|$ extracted using $\tau$ decays to strange mesons is smaller than that obtained by other methods~\cite{Nugent:2013ui}. In particular, the central value obtained using the ratio $\Gamma_{\tau\to K\nu}/\Gamma_{\tau\to\pi\nu}$ is about 1\% below the value from $\left|V_{ud}\right|$ and CKM unitarity (assuming $U_{\mu4}=0$). The value using the inclusive strange rate is even smaller, about 4\% smaller than the CKM unitarity value. While not statistically significant, these are intriguing and could be signs of a heavy neutrino component to $\nu_\tau$ since final states involving kaons have less phase space available (which is suggested by the inclusive, multibody final states leading to a smaller $\left|V_{us}\right|$). The value of $\left|U_{\tau4}\right|$ required to align the central values of $\left|V_{us}\right|$ from $\Gamma_{\tau\to K\nu}/\Gamma_{\tau\to\pi\nu}$ and CKM unitarity is in tension with the estimate of the limit from $\tau\to\mu$ decay derived in \SEC~\ref{sec:Ulimits} but, as mentioned above, there could be an unaccounted for systematic bias in this estimate. If the discrepancy in $\left|V_{us}\right|$ measurements becomes significant, it could be another hint of the existence of an ${\cal O}(100~\rm MeV)$ component to the $\tau$ neutrino.

\subsection{Matter effects on neutrino oscillations}
\label{sec:matter}
For $m_4$ below $100~\rm MeV$, the strongest limit on $\left|U_{\tau4}\right|$ is due to matter effects in atmospheric neutrino oscillations. The lack of weak interactions of the sterile neutrino leads to a difference in the matter effects between $\nu_\mu$ and the linear combination of $\nu_\tau$ and sterile that makes up the light neutrinos. Limits on $\left|U_{\tau4}\right|$ have been derived from analyzing the zenith angle distribution of muon neutrinos at Super-K~\cite{Abe:2014gda} and in IceCube and (low energy) DeepCore data in the language of neutrino NSI~\cite{Esmaili:2013fva}. The Super-K limit on $\left|U_{\tau4}\right|$ is statistics limited and will be improved with more data. An analysis of the PINGU upgrade of IceCube indicates that it will be able to place a 90\%~C.\,L. upper limit on the NSI parameter $\epsilon_{\tau\tau}$ of $1.7\times10^{-2}$~\cite{Choubey:2014iia}. This will improve the reach on $\left|U_{\tau4}\right|$ to about 0.3. Furthermore, a year of full DeepCore data will allow $\epsilon_{\tau\tau}$ to be probed at 90\%~C.\,L. to $6\times10^{-3}$~\cite{Esmaili:2013fva} which will allow values of $\left|U_{\tau4}\right|>0.2$ to be tested. This is a very promising test of the model.

\subsection{$U_{e4}$ and $U_{\mu4}$}
In addition to a nonzero $U_{\tau4}$ we might expect, at some level, that $U_{e4}$ and $U_{\mu4}$ are also not vanishing in this model. While it is technically natural for $U_{e4}$ and $U_{\mu4}$ to be extremely suppressed compared to $U_{\tau4}$ (radiative contributions to $U_{e4},U_{\mu4}$ are necessarily generated but are proportional to the light neutrino masses and are therefore tiny), it is not a requirement that the sterile-active neutrino coupling only violate $L_\tau$. (We use $L_\ell$ to label the global U(1) associated with lepton flavor $\ell$.) In most of this paper, for simplicity and in light of the phenomenological requirement that $U_{\tau4}\gg U_{e4},U_{\mu4}$, we have ignored $U_{e4}$ and $U_{\mu4}$ but it is possible that they are nonzero. In fact, it is easy to contemplate a model in which the hierarchy $U_{\tau4}\gg U_{e4},U_{\mu4}$ is enforced  with $U_{e4},U_{\mu4}\ne0$ by imposing a symmetry that satisfies minimal flavor violation (MFV). In an MFV scenario, we would expect that $\lambda_i$ in \EQ~\eqref{eq:Lm} are proportional to the lepton Yukawas so that, in addition, $U_{\mu4}\gg U_{e4}$.

If we take the reasonable view that in this model $U_{e4}$ and $U_{\mu4}$ are not strictly zero, we could potentially expect to see a signal in the observables that we used to constrain $U_{e4}$ and $U_{\mu4}$ in \SEC~\ref{sec:Ulimits}. Furthermore, we might also expect signals in lepton-flavor--violating (LFV) processes such as $\tau\to\mu\gamma$, $\tau\to e\gamma$, $\mu\to e\gamma$, or $\mu\to e$ conversion. The fact that this model includes a neutrino with a mass above $10~\rm MeV$ adds additional motivation to search for LFV--a relatively large $m_4$ reduces the GIM-suppression of such processes. The decay $\tau\to\mu\gamma$ could be particularly interesting in an MFV context, while the effort to greatly improve the reach in sensitivity to $\mu\to e\gamma$ and $\mu\to e$ conversion makes these processes interesting as well.

Lastly, a nonzero value of $U_{\mu4}$ would also open up the possibility of observing this model in $\nu_\mu\to\nu_\tau$ oscillations at the proposed short baseline experiment MINSIS~\cite{Alonso:2010wu}. MINSIS proposes to use the NUMI beamline at Fermilab with a kton-scale emulsion cloud chamber detector, capable of observing $\tau$ neutrinos, situated $1~\rm km$ away. Early studies indicate that, for $\left|U_{\tau4}\right|^2=0.1$, sensitivity to $\left|U_{\mu4}\right|^2$ above roughly $10^{-6}$ is possible~\cite{Alonso:2010wu}.


\section{Conclusions}
\label{sec:conc}
The paradigm of CDM does an excellent job of describing a wide range of data on the scales of galaxy clusters or larger. However, there appear to be persistent discrepancies between predictions in the CDM paradigm and observations at smaller scales. 

We have focused on one of these problems, that of missing satellites. This problem can be solved by introducing strong interactions between neutrinos and DM which keep the DM in thermal equilibrium with the relativistic matter in the early Universe to lower temperatures than typically expected. This washes out structures with masses below a particular scale $M_{\rm cutoff}$. If $M_{\rm cutoff}$ is chosen to be in the range $10^7-10^9M_\odot$, then the expectation for the number of satellite galaxies of a Milky Way sized galaxy can be brought into agreement with observations, solving the missing satellites problem. A cutoff of this size requires a large DM-neutrino scattering cross section. This can be realized in  a renormalizable theory if the DM has a mass that is tens of MeV and is coupled to a sterile neutrino that mixes with the active neutrinos. The strength of the mixing that is required combined with both cosmological and particle physics data implies that the sterile neutrino mixes most with $\nu_\tau$ and leads to a heavy neutrino that is mostly sterile but with a sizable $\nu_\tau$ component. There are a number of signatures of this scenario, both for astrophysical and particle physics experiments. 

Strong DM-neutrino interactions are particularly interesting for supernovae. The mass scale implied by a solution to the missing satellites problem indicates that a future observation of neutrinos from a nearby supernova could show an imprint of DM-neutrino scattering. This scenario can also be tested at neutrino oscillation experiments, due to the the change of matter effects due to the sterile neutrino. There will be progress on these measurements, probing regions of parameter space that are able to solve small scale structure problems. $\tau$ decays are also a promising area to search for the signs of neutrino-DM interactions. Improvements of the searches for a massive component of $\nu_\tau$  would be a useful way of probing this model. Lastly, lepton-flavor--violating processes are well motivated by this scenario. The reach of searches for these processes will be greatly improved in the near future, opening up the opportunity for discovery.

\acknowledgments
We thank Kfir Blum, Cora Dvorkin, Alex Fry, Akshay Ghalsasi, Oleg Gnedin, Matt McQuinn, Andrew Pontzen, Maxim Pospelov, Tom Quinn, Sanjay Reddy, Adam Ritz, and Kris Sigurdson for numerous discussions. S.I. appreciates the hospitality and stimulating atmosphere of the PACIFIC-2014 conference, supported by NSF award number PHY-1402090, where part of this work was completed. This work was supported in part by the U.S. Department of Energy under Grant No. DE-SC0011637.

\bibliography{nDM}
\bibliographystyle{jhep}
\end{document}